\documentclass[aps, nofootinbib,superscriptaddress, preprintnumbers, longbibliography]{article}
\usepackage{tikz}
\usepackage{jcappub}
\usepackage[utf8]{inputenc}
\usepackage{amsmath,amsfonts,amssymb}
\usepackage{graphicx}
\usepackage[toc,page]{appendix}
\usepackage{cleveref}
\usepackage[loose]{units}
\usepackage[markup=default]{changes}
\definechangesauthor[name={Red}, color=red]{YN}
\setdeletedmarkup{\textcolor{red}{\sout{#1}}}
\usepackage{booktabs}
\usepackage{mathtools}
\usepackage{graphicx} 
\usepackage{subcaption} 
\usepackage[margin=1in]{geometry} 
\usepackage{braket}
\usepackage[normalem]{ulem}

\begin{document}

\title{
Production of gravitational waves by inflationary transitions in aligned natural inflation
}

\author[a,b]{Federico Greco}

\affiliation[a]{Dipartimento di Fisica e Astronomia “G. Galilei”, Universit\`a degli Studi di Padova, via Marzolo 8, I-35131 Padova, Italy}
\affiliation[b]{INFN, Sezione di Padova, via Marzolo 8, I-35131 Padova, Italy}

\emailAdd{federico.greco@pd.infn.it}

\abstract{The original axion natural inflation model predicts a tensor-to-scalar ratio exceeding experimental limits. Conversely, in aligned axion inflation, inflation can proceed along trajectories emerging from near a saddle point of the two-field potential and ending through an instability in the orthogonal direction. Such solutions satisfy present observational limits and will be tested by future CMB experiments.
Previous studies have suggested the possibility of two distinct inflationary stages separated by a transition characterized by rapid oscillations of the fields. In this work, we demonstrate that the existence of these two stages is a generic feature of the model. We explore a possible phenomenological signature of the transition when a U(1) gauge field is coupled to the axions, namely, the production of gravitational waves (GWs) sourced by gauge quanta generated during the transition. This mechanism produces a feature similar to those seen in spectator axion models or axion inflation with appropriate potentials, i.e. a strongly scale-dependent power spectrum. The scale at which the GW spectrum is produced is determined by the duration of the second inflationary phase. Consequently, the spectrum may peak at different frequencies, potentially detectable by future GW experiments.
}

\maketitle

\section{Introduction}
\label{sec:intro}

The inflationary paradigm \cite{Guth:1980zm,Linde:1982uu,Albrecht:1982wi} is able to provide the initial conditions for the Universe, solving long-standing problems of the hot big-bang theory, and it is also able to explain the large-scale structure we observe at the present time through the initial quantum fluctuations. Although the success, if we assume that the initial accelerated expansion was caused by one or more fields, we still do not have a unique convincing particle physics realization of the inflationary epoch.

Restricting for the moment to the single-field case, this hypothetical field should have very peculiar properties in order to sustain a long enough period of inflation. 
Indeed, the potential of this field should be sufficiently \textit{flat}, in order to provide the correct equation of state for enough e-folds. This flatness of the potential is in general not stable under quantum corrections, especially for the so-called \textit{large field} models. In this class of models, the field spans a length in field space that is of the order of the Planck scale $M_p$.
A generic prediction of these models is an observable tensor signal \cite{Lyth:1996im}. However, any irrelevant operator of the form $\Delta V_n = c_n\, \phi^{n+4}/M^n$, where $M$ is an energy scale and $c_n$ a dimensionless coefficient generally spoils the flatness of the potential, unless $M>M_p$, which is in general a regime where we do not expect general relativity to be valid, or all the coefficients are fine-tuned to $c_n \ll 1\;,\;\; \forall n$.

A possible solution to this problem is given by a symmetry that protects the potential \cite{Freese:1990rb,Adams:1992bn}. Such symmetry can be found in axion-like particles (ALPs), which are pseudoscalar particles endowed with a shift symmetry $\phi \rightarrow \phi + c$, which is usually broken by gauge instantons to a discrete sub-group $\phi \rightarrow \phi + 2 \pi f$, with $f$ being a constant labeled axion decay constant, which allows for a non-trivial potential term for the axion. The first model to exploit the protection mechanism provided by ALPs is the so-called \textit{natural inflation} model \cite{Freese:1990rb}, which has the following potential
\begin{equation}
    V(\phi) = \Lambda^4 \left[  1- \cos \left( \frac{\phi}{f} \right) \right].
\end{equation}
Although this model is now ruled out for every value of $f$ by CMB observations \cite{Planck:2018jri}, a broad class of models known as \textit{axion inflation} has been developed in this context (see e.g. \cite{Pajer:2013fsa} for a review). Typically, such scenarios require an axion decay constant of trans-Planckian magnitude to sustain a sufficient duration of inflation. When this occurs, quantum gravity effects are expected to become relevant, leading to the violation of any continuous global symmetry, including the Peccei–Quinn symmetry associated with the axion \cite{Ghigna:1992iv,Holman:1992us,Kamionkowski:1992mf,Giddings:1987cg,Giddings:1989bq}. This issue is absent in the case of string-theoretic axions, where the shift symmetry has its origin in an underlying gauge symmetry and is therefore protected from quantum gravity corrections \cite{Kallosh:1995hi}. However, explicit and fully controlled string theory constructions that realize trans-Planckian decay constants are still unknown \cite{Banks:2003sx}.  
Several proposals have been put forward to address this limitation. Among the most studied are \textit{N-flation} \cite{Dimopoulos:2005ac}, where inflation arises from the collective behavior of multiple sub-Planckian axions, \textit{axion monodromy} \cite{McAllister:2008hb}, in which a monodromic structure effectively enlarges the field range of a string axion beyond the Planck scale, and   \textit{aligned natural inflation} \cite{Kim:2004rp}, which is the focus of this work. The mechanism at the basis of this model involves two axionic degrees of freedom subject to a potential that selects an almost flat combination of the two fields. This nearly flat direction acts as an effective field with a large decay constant, even though each of the fundamental axions remains sub-Planckian.

The alignment mechanism proposed in~\cite{Kim:2004rp} has attracted considerable attention, leading to a number of follow-up studies and concrete realizations aimed at achieving compatibility with quantum gravity. It was shown in~\cite{Rudelius:2015xta,Montero:2015ofa} that the large effective axion decay constant that appears in~\cite{Kim:2004rp} can be understood more explicitly by choosing a different field basis in which the potential becomes diagonal, and the enhancement arises as an eigenvalue of the non-diagonal kinetic matrix.  
In addition, multi-field extensions of axion inflation have been widely explored~\cite{Rudelius:2015xta,Montero:2015ofa,Brown:2015iha,Bachlechner:2015qja,Hebecker:2015rya,Brown:2015lia,Heidenreich:2015wga,Kappl:2015esy,Choi:2015aem}, particularly in the context of the \textit{weak gravity conjecture} (WGC)~\cite{Arkani-Hamed:2006emk,Cheung:2014vva}, which is expected to hold in any consistent theory of quantum gravity. These analyses indicate that the minimal version of the alignment mechanism~\cite{Kim:2004rp} does not satisfy the WGC. However, additional instanton contributions, although subdominant in the scalar potential, can introduce additional interactions, potentially making the mechanism consistent with the conjecture~\cite{Bachlechner:2015qja,Hebecker:2015rya,Brown:2015iha,Brown:2015lia}.  
Such corrections may also induce small modulations in the potential, which may leave imprints on the phenomenology of CMB~\cite{Abe:2014xja,Kappl:2015esy,Choi:2015aem}. The embedding of the alignment scenario within string theory has been further investigated in~\cite{Long:2014dta,Gao:2014uha,Hebecker:2018fln,Palti:2015xra,Angus:2021jpr}.~\footnote{Additional instanton effects can also give rise to the multi-natural inflation mechanism discussed in~\cite{Czerny:2014wza,Daido:2017wwb}.}

Beyond the issue of embedding the model in a full quantum gravity setting, a number of works have studied the phenomenology associated to aligned natural inflation which turned out to be more complex than what was initially supposed. In \cite{Peloso:2015dsa}, it was shown that there are two different kinds of inflationary trajectories associated with the model: a first class (\textit{stable}), made of inflationary paths which terminate at the minima of the potential and a second class where the end of inflation is caused by a tachyonic instability in the heavy direction, which no longer stabilizes the valley. We call these second class of trajectories \textit{metastable}.
The phenomenology of the first kind of solutions is ruled out since it predicts a tensor signal greater than that of natural inflation. In contrast, the metastable paths are compatible with the CMB constraints because the inflationary trajectory takes place near a saddle point, where the potential is flatter and leads to a suppressed tensor signal. In \cite{Peloso:2015dsa}, the analysis of the different paths was performed mostly numerically, while in \cite{Greco:2024ngr} an analytical study of the metastable phase was carried out, leading to simple equations that control the parameter space of the model and the associated phenomenology.

Another feature presented in \cite{Greco:2024ngr} is the possibility of having transitions between metastable and stable trajectories characterized by an oscillatory behavior of the fields. Here, we point out that the presence of couplings of the axions to one or more gauge fields can lead to an important production of scalar and tensor perturbations.
Therefore, in this work, we consider the aligned natural inflation model coupled to a U(1) gauge field and we focus on the production of gravitational waves (GWs) sourced by gauge quanta which in turn are sourced by the oscillating behavior of the axion fields during the transition.

We denote the two axionic fields of the model as $\theta$ and $\rho$. The mass eigenstates consist of linear combinations of these two fields, found by diagonalizing the mass matrix. The alignment condition produces a hierarchy between these two mass eigentstates, so that we can define the heavy, $\psi_H$, and light , $\psi_L$, states. The first phase of inflation terminates due to a tachyonic instability in $\psi_H$, which no longer stabilizes the trajectory. At this point, the system is attracted towards a new region, where $\psi_H$ is stable. Here, the fields undergo damped oscillations about the stable heavy direction. After this stage, the second inflationary period starts as the fields move along the light direction connected to the minimum.

After the tachyonic instability and the end of the second phase of inflation, we have $\psi_H \simeq \theta$. Therefore, if the U(1) field coupling to $\rho$ is negligible, the gauge field production is significant only during the transition, because the damped oscillations are performed mostly along the $\theta$ direction. On the contrary, after the system has settled in the new stable trajectory towards the minimum of the potential, the coupling of the gauge field to $\theta$ does not lead to a significant production of gauge quanta.

Therefore, for simplicity, in this work, we restrict our attention to the coupling between the gauge field and $\theta$.
This simplification can be motivated by scenarios in which the axion–gauge interaction is induced through loop effects of ``matter'' fields carrying charges under both the confining gauge group, responsible for generating the axion potential, and the U(1) gauge group. In such cases, the heavier axionic field is expected to exhibit a stronger coupling to the U(1) field, which in our setup corresponds to $\theta$.

In this work, we study analytically the second phase of inflation, and we are thus able to estimate the number of e-folds after the transition for a given set of initial parameters, showing that the presence of two distinct stages of inflation is a generic feature of the metastable solutions. Moreover, we perform a numerical study of the coupled dynamics of the system taking into account the backreaction of the gauge field on the equations of motion of the two axions. We find a considerable enhancement of the GW power spectrum, potentially observable by future experiments, with a characteristic behavior reflecting the oscillatory nature of the transition. We performed the analysis only for two specific examples, namely for specific choices of the parameters of the model, as the scope of the paper is to show the possible phenomenological signal of the transitions in aligned natural inflation. However, a similar phenomenology can be expected in models which predict the existence of two or more separated inflationary stages.

This work is organized as follows. In Section \ref{sec:model}, we describe the aligned natural inflation model coupled to a U(1) gauge field. We then discuss the equations of motion that govern the dynamics of the system and provide an analytic estimate of the duration of the second inflationary stage. In Section \ref{sec:tensor_pert}, we derive the action for the tensor perturbations sourced by the gauge field and compute their power spectrum using the Green’s function method. In Section \ref{sec:numerical}, we explain the numerical approach employed to evaluate the GW power spectrum. More details are given in Appendices \ref{App:Code_variables} and \ref{app:cut-off}. Finally, we summarize our findings and conclude in Section \ref{sec:conclusion}.

\section{Aligned axion inflation coupled to a U(1) gauge field}
\label{sec:model}

\subsection{The model}
\label{subsec:model}

The aligned axion inflation model coupled to a U(1) gauge field is defined in terms of two axion-like fields coupled through a topological term to the gauge field. The action is given by ~\footnote{We use natural units and assume a spatially flat FLRW spacetime, $ds^2 = -dt^2 + a^2(t)\, \delta_{ij} \,dx^i\,dx^j$, where $a(t)$ is the scale factor. The reduced Planck mass and the Hubble rate are defined as $M_p = (8\pi G)^{-1/2}$ and $H = \dot{a}/a$, respectively, with dots denoting derivatives with respect to cosmic time $t$.
}
\begin{equation}
{\cal S} = \int {\rm{d}} ^4 x \sqrt{-g} \left[\frac{M_p^2}{2} R  - \frac{1}{2} \left( \partial \theta \right)^2  - \frac{1}{2} \left( \partial \rho \right)^2 
- V(\theta, \rho) -\frac{1}{4} F^2 -\frac{1}{4} \left( \frac{\theta}{F}  + \frac{\rho}{G} \right) F \tilde{F}    \right],
\label{action}
\end{equation}
where $F_{\mu \nu} = \partial_\mu A_\nu - \partial_\nu A_\mu$ and $\tilde{F}^{\mu \nu} =\frac{ \epsilon^{\mu \nu \rho \sigma} F_{\rho \sigma}}{2 \sqrt{-g}} $ is the Hodge dual of the field strength 2-form $F$ and $\epsilon$ is the completely antisymmetric tensor defined by $\epsilon^{0123}=+1$. 
The potential for the axion fields is given by
\begin{equation}
 V(\theta, \rho)= \sum_{i=1,2} \Lambda_i^4 \left[ 1 - \cos \left( \frac{\theta}{f_i} + \frac{\rho}{g_i} \right) \right].  
\end{equation}
We assume $\Lambda_1 \geq \Lambda_2$ without loss of generality, and we also define the following parametrization
\begin{equation}
\Lambda^4 := \Lambda_1^4 + \Lambda_2^4 \;\;,\;\; r_\Lambda := \frac{\Lambda_2^4}{\Lambda_1^4} \;. 
\label{Lambda-resca}
\end{equation}
where the request for $\Lambda_1 \geq \Lambda_2$ translates into $r_\Lambda \leq 1$.
The model has a global internal symmetry given by the combined rotations
\begin{eqnarray}
\left( \begin{array}{c}
\theta \\  \rho 
\end{array} \right) \;\to\;
 R
\left( \begin{array}{c}
 \theta \\  \rho 
\end{array} \right) \;\;,\;\; 
\left( \begin{array}{c}
f_i^{-1} \\ g_i^{-1} 
\end{array} \right) \;\to\;
 R
\left( \begin{array}{c}
f_i^{-1} \\ g_i^{-1} 
\end{array} \right) \;\;,\;\;  
\left( \begin{array}{c}
{F} \\ {G} 
\end{array} \right) \;\to\;
 R
\left( \begin{array}{c}
{F} \\ {G} 
\end{array} \right) \;\;,\;\;
i = 1,\,2 \;,
\label{rotation}
\end{eqnarray}
where $R \in$ SO(2) is a generic global rotation matrix. The physical observables have to be invariant under this global transformation. In~\cite{Greco:2024ngr}, the model has been studied in the absence of the U(1) gauge field (or equivalently, in the limit of large $F$ and $G$ in~\eqref{action}) and the phenomenology has been written in terms of rotational invariant parameters. We consider here the same invariant parameters with the addition of a new parameter, defining the coupling to gauge fields. The invariant parameters are
\begin{equation}
n_1 := f_1^{-2} + g_1^{-2} \;\;,\;\; n_2 := f_2^{-2} + g_2^{-2} \;\;,\;\;  n_3 := F^{-2} + G^{-2} \;\;,\;\; {\cal C} := f_2^{-1} g_1^{-1} - f_1^{-1} g_2^{-1} \;.
\label{NC}
\end{equation}
These parameters are defined in terms of scalar and vector products of the row vectors defined in Eq. (\ref{rotation}). We can exploit the symmetry redundancy by eliminating one of the axion decay constants or one of the couplings with the gauge fields, therefore we choose to ``fix the gauge" by setting $g_2^{-1}=0$. 
The ``aligned" in the name of the model refers to the following condition
\begin{equation}
    \frac{f_1}{g_1} = \frac{f_2}{g_2},
    \label{eq:alignment-condition}
\end{equation}
which corresponds to ${\cal{C}}=0$. When this condition is met, the potential develops a flat direction which can sustain inflation. Condition (\ref{eq:alignment-condition}) has to hold at an approximate level in order to have an almost flat direction. This alignment can be accidental, or due to a symmetry in the mechanism responsible for the breaking of the two shift symmetries $\theta \rightarrow \theta + c$, $\rho \rightarrow \rho + c$. In the second case, a small breaking of this symmetry will ensure a sufficient alignment.

In the following, we parameterize the degree of alignment through the dimensionless parameter
\begin{equation}
    \gamma := \frac{2 \; {\cal{C}}}{n_1+n_2},
    \label{eq:gamma-definition}
\end{equation}
and, since $\gamma$ is dimensionless, while $n_i$ and ${\cal C}$ have mass dimension $-2$, in the following computations, every expansion in small ${\cal{C}}$ is, more properly, an expansion in $\left\vert \gamma \right\vert \ll 1$. 

The model possesses two different classes of inflationary solutions \cite{Peloso:2015dsa}. One is given by trajectories in field space where the end of inflation corresponds to the fields setting at the vacuum of the potential, while the other one is given by trajectories starting at a saddle point of the potential where the end of inflation is given by a tachyonic instability in the direction orthogonal to the trajectory of the fields. For a discussion on these trajectories and the condition that the parameters must satisfy for their existence, see \cite{Greco:2024ngr}.
The phenomenology of the first class of solutions is ruled out by the CMB bounds on the tensor-to-scalar ratio \cite{Tristram:2021tvh,Galloni:2022mok}, while the second class of solutions is still viable due to the suppression of the tensor signal that it exhibits
~\cite{Peloso:2015dsa}.

\subsection{Analytical treatment of the second inflationary stage}
\label{subsec:Second-inflation}
Another feature of the model is the presence of two distinct phases of inflation. This is possible because at the end of the first metastable inflationary trajectory, after the tachyionic instability, the fields reach a valley belonging to the first class, therefore leading to a second phase of inflation which terminates in the minimum of the potential. If this second phase lasts less than about $50-60$ e-folds, then the CMB phenomenology is not affected much and we can rely on the analytic results already obtained for the second class of solutions. However, the transition between the two phases is characterized by a short fast oscillatory behavior of the fields as shown in Fig ~\ref{fig:traj}.
The existence of solutions characterized by both phases of inflation was found in \cite{Greco:2024ngr}, but it was not explored whether this was a generic behavior of the model, or just a peculiar feature found in those sets of parameters.

Here, although we are unable to follow analytically the precise behavior of the transient phase, we can provide an analytical estimate of the number of e-folds that the system spends in the second inflationary stage for any given set of parameters of the model. To do so, we diagonalize the mass matrix at the origin, finding the light direction which corresponds to the valley along which we have the second phase of inflation. We introduce the canonically normalized field $\varphi$ to parameterize the valley, finding the expressions for $\theta (\varphi)$ and $\rho(\varphi)$. All the details of these calculations are given in Appendix \ref{App:second-inflation}. We find the effective Lagrangian along the second inflationary valley, given by
\begin{equation}
    {\cal{L}}_{\rm{eff}}(\varphi, \dot{\varphi})=\frac{1}{2} \dot{\varphi}^2-V_{\rm{eff}}(\varphi),
\end{equation}
with
\begin{equation}
    V_{\rm{eff}}(\varphi)= \frac{\Lambda^4}{2(1+r_\Lambda)}\left[ \frac{r_\Lambda {\cal{C}}^2}{n_1+n_2 r_\Lambda} + {\cal{O}}({\cal{C}}^4)  \right] \varphi^2 + {\cal{O}}(\varphi^4).
    \label{eq:Veff}
\end{equation}
where we have expanded the potential to second order in $\varphi$ and ${\cal{C}}$. We notice, as expected, that in the limit of perfect alignment, ${\cal{C}}=0$, the valley becomes exactly flat. 
Using the slow-roll approximation, we can analytically solve the equations of motion finding the number of e-folds in this second stage (that we denote as $N_{\rm extra}$) as a  function of the value assumed by $\varphi$ (that we denote as $\varphi_0$) at the start of this second stage:
\begin{equation}    
N_{\rm extra}
=\frac{\varphi_0^2}{4 M_p^2}- \frac{1}{2}.
\label{eq:e_folds_analytic}
\end{equation}
The initial value of $\varphi$ is estimated using the equation
\begin{equation}
    \rho_{\rm{end}}=\rho(\varphi_0),
    \label{eq:phi0}
\end{equation}
where $\rho_{\rm{end}}$ is the value of $\rho$ at the end of the first phase of inflation. This estimate relies on the fact that the field $\rho$ is nearly stationary during the transition (see fig. (\ref{fig:traj})).
In the right hand side of this equation we insert the rotation that relates $\varphi$ to the original fields (along the second inflationary valley, namely setting the heavy field to zero), this gives eq. (\ref{eq:parametrized_theta_rho}). For the left hand side, we use the analytic approximation for $\rho_{\rm{end}}$ derived in \cite{Greco:2024ngr} and here reported in eq.(\ref{eq:rho_end_approx}). 

In this way, the relation~\eqref{eq:phi0} becomes an equation in which the only unknown is $\varphi_0$. We solve this equation, and then insert the value of $\varphi_0$ in eq.~\eqref{eq:e_folds_analytic} to determine the estimate for the number of e-folds in this second phase of inflation.
 We have compared these analytic results with the numerical evolution of the exact equations which lead to the two inflationary stages, and indeed we find that this procedure provides a good estimate of the total number of e-folds of the second phase of inflation, with a relative error of about $10 \%$. 
 
\begin{figure}
\centering
\includegraphics[scale=0.45]{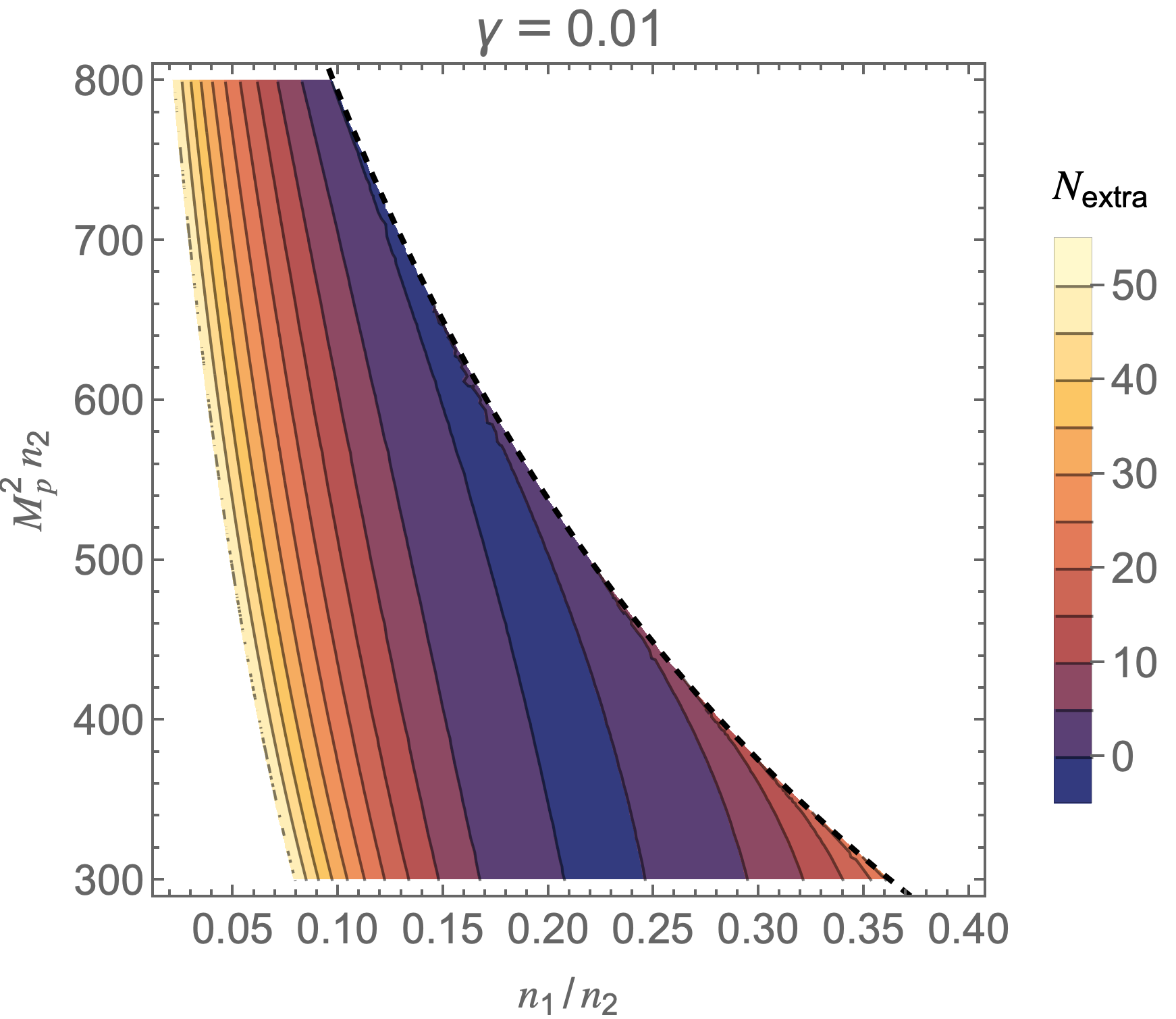}
\includegraphics[scale=0.45]{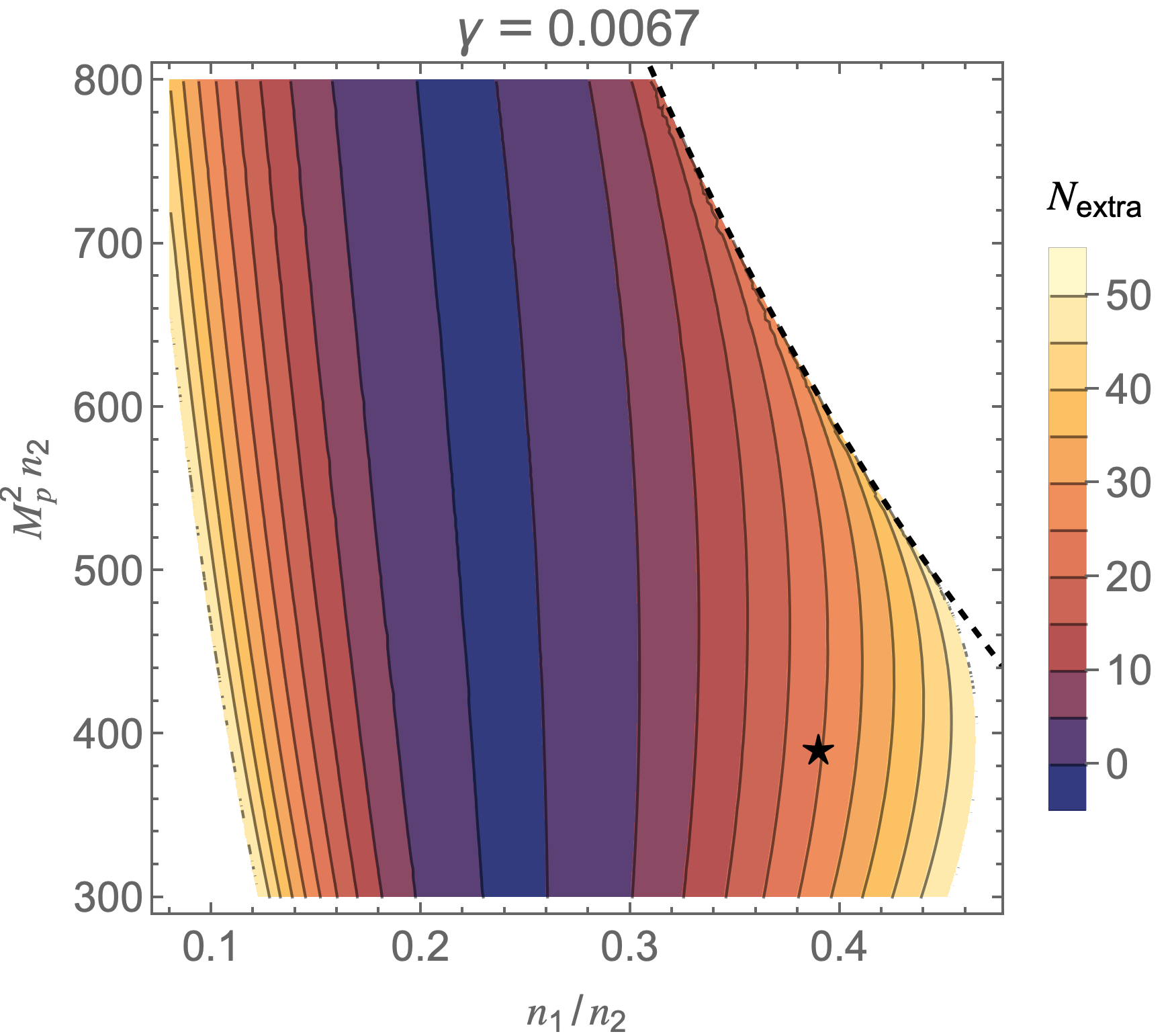}
\caption{Contour plots of the number of e-folds obtained in the second phase of inflation as a function of the parameters of the model and for two different values of alignment $\gamma$. The black dashed line is given by $\tilde{n}_2=\tilde{n}_{2,{\rm{max}}}(r_1)$ while the black star on the right panel marks the parameters used to compute the power spectrum in the left panel of fig. (\ref{fig:power_spectrum}).}
\label{fig:e-folds-extra}
\end{figure}

 Since the parameter space of the model is composed of $\{ n_1,n_2,r_\Lambda,{\cal{C}} \}$, to visualize how many e-folds of second inflation we generally obtain, we fix the value of the scalar tilt, found considering only the first phase of inflation, to the preferred value of $n_s=0.965$ \cite{Planck:2018jri}. By doing so, we are able to express the phenomenology of the model in terms of two parameters at fixed value of the alignment $\gamma$ \cite{Greco:2024ngr}
\begin{equation}
    r_1 := \frac{n_1}{n_2},\;\; \tilde{n}_2 := M_p^2\, n_2,
    \label{eq:phenomenoloy-parameters}
\end{equation}
with
\begin{equation}
    \tilde{n}_2 \,\le \,\frac{1-\sqrt{r_1}}{q(1+r_1)} := \tilde{n}_{2,{\rm{max}}}(r_1),\;\; q := \frac{\gamma^2}{4\,(1-n_s)}.
\end{equation}
The other two parameters can be expressed as functions of $r_1$ and $\tilde{n}_2$ as
\begin{equation}
    {\cal{C}}= \frac{\gamma \, (1+r_1)\,\, \tilde{n}_2}{2 \,M_p^2},\;\; r_\Lambda= \frac{r_1}{1-q\,\,(1+r_1)^2 \,\,\tilde{n}_2}.
\end{equation}
We can use $\{r_1, \tilde{n}_2 \}$ to study the number of extra e-folds in the model. In fig. (\ref{fig:e-folds-extra}), we show a contour plot of the number of second inflationary e-folds as a function of these parameters for two different values of the alignment $\gamma$. We show the contour lines up to $N=50$ e-folds, keeping in mind, however, that these extra e-folds slightly modify the predicted scalar tilt affecting also the CMB phenomenology. In fact, when we fix $n_s$ in the above equations, we are not considering the second stage of inflation. The presence of these extra e-folds slightly changes the point in the metastable trajectory where the CMB modes are produced. Therefore, the physical $n_s$ observed is not $n_s=0.965$, but depends on the duration of the second inflationary period and, therefore, is slightly modified \footnote{Fixing the value of $n_s$ serves merely as a way to illustrate the number of additional e-folds. We adopt the preferred value since the inclusion of extra e-folds modifies it only marginally.}.  

As we can see, we can generically obtain a prolonged second inflationary stage. Therefore, the phenomenology that we study in this paper is not restricted to a small portion of the parameter space of the model, on the contrary, the presence of both classes of inflationary trajectories is a generic feature of the aligned axion inflation model.

\subsection{Equations of motion}
\label{subsec:equations-of-motion}
The period of transition between the two inflationary stages is a source of gravitational waves produced by the gauge field excited by the fast oscillations of the axions, as we describe successively.
\begin{figure}
\centering
\includegraphics[scale=0.8]{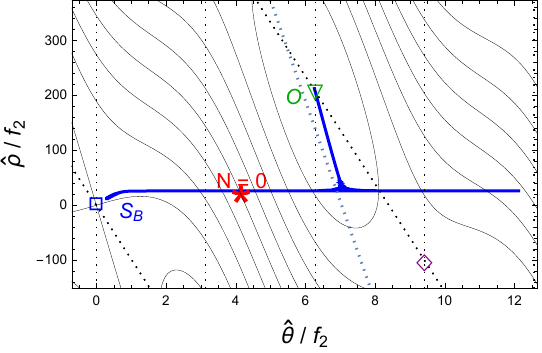}
\includegraphics[scale=0.8]{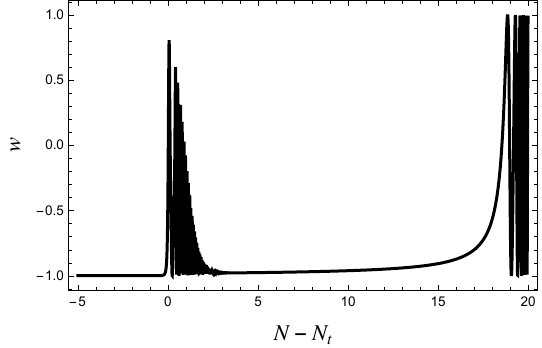}
\caption{Field trajectory (left panel) and equation of state as a function of the number of e-folds (right panel) for $n_1 = \frac{70}{M_p^2} ,\, n_2 = \frac{500}{M_p^2} ,\, \gamma = 0.0067 ,\, r_\Lambda = 0.2$. 
In this figure the fields $\hat \theta$, $\hat \rho$ are shifted with respect to the original fields so that the saddle point $S_B$ is at the origin in these coordinates. $N=0$ marks the end of the first phase of inflation which is followed by oscillations about a new inflationary valley connected to the minimum ${\cal O}$, and by a second inflationary stage of duration $N_{\rm extra} \simeq 18$ along this valley, and, finally, by oscillations about the minimum. In this figure, the backreaction of the gauge field is neglected.
}
\label{fig:traj}
\end{figure}

As explained in the introduction, we consider the case in which the gauge field is only coupled to $\theta$, setting $G^{-1} = 0$.
This can be a reasonable simplification if the axion–gauge interactions are generated through loop effects of ``matter'' fields charged under both the confining gauge group (responsible for the axion potential) and the U(1) group, in strong analogy to the typical mechanism in which the QCD axion–photon coupling is induced from loops of particles charged both under SU(3) color and U(1) electromagnetic (see e.g.~\cite{DiLuzio:2020wdo} for a review). In this case, it is natural to expect that the axion predominantly coupled to the confining sector, and hence acquiring the larger mass, also exhibits the stronger coupling to the U(1) group. We refer here to the physical particle spectrum, which is the one found at the true vacuum and therefore at the minimum of the potential.  The heavy eigenstate is a linear combination of the fields given by $\psi_H = v_{h,1} \theta + v_{h,2} \rho $, where the expressions for $v_{h,i}$ are given in eq. (\ref{eq:heavy-eigenvector}). For $\gamma \ll 1$, we have
\begin{equation}
    v_{h,1} \simeq 1+ {\cal{O}}(\gamma ^2),
    \label{eq:heavy_eigenvec}
\end{equation}
\begin{equation}
    v_{h,2} \simeq {\cal{O}}(\gamma ^2),
    \label{eq:light_eigenvec}
\end{equation}
therefore, we have $\psi_H \simeq \theta$ at the minimum of the potential.
Since $\theta$ almost coincides with the heavy eigenstate of the theory, we expect $F^{-1} \gg G^{-1}$. In any case, the coupling with $\theta$ is the most important source of GWs at the transition, since the oscillations, as one can see in fig. \ref{fig:traj}, take place along $\theta$, leading to $\dot{\theta} \gg \dot{\rho}$ and, as we explain later in this section, the time derivative of the fields, together with the couplings, controls the production of gauge quanta.
However, this no longer holds after the transition, since during the second stage of inflation we have $\dot{\rho} \gg \dot{\theta}$, and for comparable couplings this phase would dominate the production of gauge quanta.
This behavior is typical of axion inflation models coupled to gauge fields. In most cases, most of the GW production occurs toward the end of inflation or during preheating, when the axion velocity is maximal. This, in turn, leads to strong constraints on the axion–gauge coupling \cite{Adshead:2019lbr}, which, however, do not apply in our setup if the gauge field is only negligibly coupled to the light direction $\rho$.

We now turn to the study of the dynamics of the model. In order to perform the canonical quantization procedure, we fix the gauge adopting the radiation gauge in vacuum for $A_\mu$, namely $A_0=\partial_i A_i=0$ (both relations can be satisfied as long as the axion coupled to the gauge field can be described as homogeneous). Therefore, the equations of motion can be written as
\begin{align}
    &\theta'' + 2 {\cal{H}} \theta' + a^2 \partial_{\theta} V - \frac{a^2}{F}  \vec{E} \cdot \vec{B}=0 , \label{eq:theta} \\
    &\rho'' + 2 {\cal{H}} \rho' + a^2\partial_{\rho} V - \frac{a^2}{G}  \vec{E} \cdot \vec{B}=0, \label{eq:rho} \\
    &{\cal{H}}' + \frac{1}{6 M_p^2} \left( \theta'^2 + \rho'^2 \right) - \frac{2 a^2 V}{3 M_p^2} + {\cal{H}}^2 = 0, \label{eq:H} \\
    &A_i'' - \partial^2 A_i - \left( \frac{\theta ' }{F}+ \frac{\rho ' }{G} \right) \epsilon_{ijl} \partial_j A_l = 0, \label{eq:Ai}
\end{align}
where the prime denotes derivatives with respect to conformal time and \({ \cal{H}}= a H = a'/a \) is the comoving Hubble parameter, while the electric and magnetic fields are the classical versions of the operators defined in eq.(\ref{eq:electric:magnetic_definition}).
The equations of motion derived above are valid under the assumption that the axionic fields remain homogeneous. However, the backreaction of the gauge field on the axion dynamics could violate this condition, potentially invalidating the approximation adopted in this work. Assessing the validity of this assumption requires a computation of the scalar perturbations of the system, which we leave for future studies. Consequently, the present work should be regarded as a first step toward the identification of the possible phenomenological signatures of aligned natural inflation, while keeping in mind that the homogeneity assumption for the axion fields must be revisited in future analyses.
We decompose the gauge field in Fourier modes and circular polarizations, and we quantize it canonically, obtaining the corresponding operator
\begin{equation}
    \label{espansione A}
    \hat{A}_i(\tau ,\vec{x})=  \int \frac{d^3 k}{(2 \pi)^{\frac{3}{2}}} \sum_{\lambda = \pm} \left[\epsilon_{\lambda,i}(\vec{k})A_{\lambda} (\tau,\vec{k}) \hat{a}_{\lambda}(\vec{k}) e^{i \vec{k}\cdot \vec{x}}+ \text{H.c.} \right],
\end{equation}
where the basis vectors satisfy the following relations
\begin{equation}
    \vec{k} \cdot \vec{\epsilon}_{\pm}(\vec{k}) = 0,
\end{equation}
\begin{equation}
    \vec{k} \times \vec{\epsilon}_{\pm} (\vec{k}) = \mp i |k| \vec{\epsilon}_{\pm}(\vec{k}),
\end{equation}
\begin{equation}
    \vec{\epsilon}_\lambda (\vec{k}) ^* = \vec{\epsilon}_\lambda (-\vec{k}) = \vec{\epsilon} _{-\lambda} (\vec{k}), 
\end{equation}
and the $+\;(-)$ sign correspond to left (right)-handed polarizations. The creation and annihilation operators satisfy the canonical commutation relations $[a_\lambda(\vec{k}),a^{\dagger}_\rho(\vec{p})]= \delta_{\lambda \rho} \delta^{(3)}(\vec{k}-\vec{p})$ and they act on the Fock space defined by
\begin{equation}
    \hat{a}_\lambda(\vec{k}) \ket{0}_{\rm{BD}}=0, \;\; \forall\,  \vec k \in \mathbb{R}^3, \,\lambda=+,\,-,
\end{equation}
where $\ket{0}_{\rm{BD}}$ is the Bunch-Davies vacuum, namely the asymptotic vacuum state for $\tau \rightarrow - \infty$, where the metric can be approximated by a Minkowski spacetime.
Using the equation of motion (\ref{eq:Ai}), we obtain the equation obeyed by the mode functions of the gauge field
\begin{equation}
    A''_{\pm}(\tau,k) + \left[ k^2 \mp k\left(  \frac{\theta'}{F}+ \frac{\rho'}{G} \right)  \right] A_{\pm}(\tau, k)=0,
    \label{eom_gauge}
\end{equation}
which is often written as
\begin{equation}
A_\lambda''(\tau,k) + \left[ k^2 - \lambda \, 2 k \xi \, a H  \right] A_\lambda(\tau,k) = 0,
\label{eq:A:xi}
\end{equation}
having defined the quantity $\xi$
\begin{eqnarray}
\xi := \frac{\dot{\theta}}{2 F H}+\frac{\dot{\rho}}{2 G H} = \frac{1}{2 a H}\left(  \frac{\theta'}{F}+ \frac{\rho'}{G}  \right).
\end{eqnarray}

\begin{figure}
\centering
\includegraphics[scale=1]{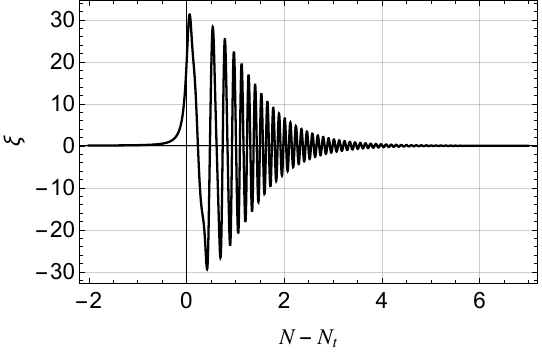}
\caption{Plot of the background behavior of \(\xi(N-N_t)\), where $N$ is the number of e-folds and $N_t$ is the value of $N$ at the end of the first inflationary phase. We notice the oscillations caused by the transition between the two inflationary stages lead to the production of both polarizations. This run has been realized with the same set of parameters of fig. (\ref{fig:traj}) without including the backreaction of the gauge field on the axions.}
\label{fig:xi_background}
\end{figure}

From eq.(\ref{eq:A:xi}), we notice that the coupling to the axion fields modifies the dispersion relation of the gauge field. When $k^2- \lambda 2 k \xi a H<0$, the mode $k$ becomes tachyonic and we have a production of quanta with comoving momentum $k$. Furthermore, we notice that the sign of the produced polarization depends on the sign of $\xi$. 
Usually in the literature on gauge field amplification during inflation, only one polarization is considered, while the other is disregarded. This is justified by the fact the the speed of the axion coupled to the gauge field is generically monotonic during inflation. Therefore, only one polarization is effectively excited, while the other can be safely neglected.
In the case of de Sitter geometry and constant $\xi$, the gauge field amplification is exponentially sensitive to $\xi$ and one can obtain an exact solution of eq.(\ref{eq:A:xi}) ~\cite{Anber:2009ua}. In the general case, $\xi$ is a time-dependent quantity, but still remains the most immediate parameter to quantify the amount of gauge field production.
In our case, the oscillations that the fields perform at the transition between the two inflationary stages cause $\xi$ to oscillate and to assume both positive and negative values, as shown in fig. (\ref{fig:xi_background}). Therefore, we also have a production of negative helicity states, thus, in principle, we should take into account both polarizations in the calculations.
In practice, as discussed in Section \ref{sec:results}, we still have that the energy density in the negative helicity modes is (for the most produced ones) at least one order of magnitude smaller than that of the positive helicity modes. 
This leads to a negligible contribution in the final computation of the GW power spectrum\footnote{This is not true for frequencies higher than the transition scale. This is explained in Section \ref{sec:results}.}. 
The ratio between the power spectra computed with two and one polarization is ${\cal O}(1)$, while the computational time is approximately four times longer. Therefore, we decided to take into account only one polarization in the final computation of the power spectrum. However, we always write down the equations considering both polarizations for the sake of generality.

Although we do not necessarily assume that the U(1) gauge field corresponds to the Standard Model photon, for simplicity we define the associated electric and magnetic field operators in analogy with those of electromagnetism
\begin{equation}
    \hat{E}_i(\tau, \vec{x}) = -\frac{1}{a^2} \hat{A}'_i(\tau, \vec{x}), \;\;\; \hat{B}_i(\tau, \vec{x})= \frac{1}{a^2}\epsilon_{ijk}  \partial_j \hat{A}_k(\tau, \vec{x}).
    \label{eq:electric:magnetic_definition}
\end{equation}
Following \cite{Garcia-Bellido:2023ser}, we can express in a compact way the electric and magnetic fields in terms of the mode functions. To this end, we define
\begin{equation}
\hat{C}^E_i := \hat{E}_i(\tau, \vec{x}) , \quad \hat{C}^B_i := \hat{B}_i(\tau, \vec{x}),
\end{equation}
from which we have
\begin{equation}
\hat{C}^\alpha_i(\vec{x}, \tau) = \frac{1}{a(\tau)^2} \int \frac{{\rm{d}}^3k}{(2\pi)^{3/2}} \sum_{\lambda= \pm}  
\left[ e^{i\vec{k}\cdot\vec{x}} \epsilon_{i,\lambda}(\vec{k}) F_{\alpha,\lambda}(\tau, k) \, \hat{a}_\lambda (\vec{k}) +  e^{-i\vec{k}\cdot\vec{x}} \epsilon_{i,\lambda}(-\vec{k}) F_{\alpha,\lambda}^*(\tau, k) \, \hat{a}_\lambda ^\dagger(\vec{k}) \right],
\end{equation}
where $\alpha= \{ E,B \}$ and the mode functions are given by
\begin{equation}
F_{\alpha,\lambda}(\tau, k) = \big\{ -A_\lambda '(\tau, k), \, k A_\lambda (\tau, k)\;  \big\}.
\end{equation}
In the equations of motion for the axions, we therefore substitute $\vec{E}\cdot\vec{B}$ with the expectation value of this operator on the Bunch-Davies vacuum state, given by the so-called \textit{backreaction integral}
\begin{equation}
     \left\langle \hat E_i \cdot \hat B_i \right\rangle =  -\frac{1}{4 \pi^2 a^4} \int {\rm{d}}k \; k^3 \frac{\partial}{\partial \tau} \left[ |A_+|^2 -|A_-|^2 \right].
     \label{eq:backreaction_integral}
\end{equation}

\section{Generation of tensor perturbations}
\label{sec:tensor_pert}
In this section, we derive the dynamics of tensor perturbations following \cite{Garcia-Bellido:2023ser}, while extending the analysis to include both polarizations of the gauge field.
To obtain the action for the tensor perturbations, we decompose the line element into the background plus perturbations as
\begin{equation}
    ds^2 = a^2(\tau) \left[ -d \tau^2+ \left( \delta_{ij} +\hat{h}_{ij}(\tau,\vec{x}) \right) dx^i dx^j \right],
\end{equation}
\noindent with $ \hat{h}_{ii}=0$ and $\partial_i\hat{h}_{ij}=0$. Expanding the action at quadratic order around the homogeneous background, one finds the following expression
\begin{equation}
    S_{\text{GW}}= \int {\rm{d}}^4 x \left[   \frac{M_p^2 a^2}{8} \left( \hat{h}^{\prime 2}_{ij} - \partial_k \hat{h}_{ij}^2 \right) + \frac{a^2}{2} \hat{h}_{ij}  \hat{T}_{ij}^{{\rm{TT}}} \right],
    \label{eq:action_GW}
\end{equation}

\noindent where $\hat{T}_{ij}^{{\rm{TT}}}$ is the transverse traceless part of the spatial components of the energy momentum tensor of the gauge field, given by
\begin{equation}
    \hat{T}_{ij}^{{\rm{TT}}}= -a^2 \left( \hat{E}_i \hat{E}_j+ \hat{B}_i \hat{B}_j
     \right)^{\rm{TT}}.
\end{equation}
We decompose the tensor perturbation in the basis of circular polarization in momentum space
\begin{eqnarray}
    \hat{h}_{ij} (\tau, \vec{x}) &&= \int \frac{{\rm{d}}^3 k}{(2 \pi) ^{3/2}} \;\; {\rm {e}}^{i \vec{k} \cdot \vec{x}}\;\; \sum_{\lambda} \Pi^*_{ij,\lambda} (\hat{k}) \;\;\hat{h}_{\lambda}(\tau,\vec{k})\nonumber\\
    &&:= \frac{2}{M_p a(\tau)}\int \frac{{\rm{d}}^3 k}{(2 \pi) ^{3/2}} \;\; {\rm {e}}^{i \vec{k} \cdot \vec{x}}\;\; \sum_{\lambda} \Pi^*_{ij,\lambda} (\hat{k})\;\; \hat{Q}_{\lambda}(\tau,\vec{k}),
\end{eqnarray}
\noindent where we have defined the polarization operators $\Pi^*_{ij,\lambda}(\hat{k}) := \epsilon_{i,\lambda}(\hat{k})\; \epsilon_{j,\lambda}(\hat{k})$ and we have introduced the operator $\hat{Q}_{\lambda}(\tau,\vec{k})$ in order to have a canonical kinetic term. Using the action given in eq.(\ref{eq:action_GW}), we derive the equation of motion for $\hat{Q}$, which is
\begin{equation}
    \left( \frac{\partial^2}{\partial \tau^2}+k^2-\frac{a''}{a} \right) \hat{Q}_\lambda (\vec{k},\tau)=- \frac{a^3}{M_p} \Pi_{ij,\lambda} (\hat{k}) \int \frac{{\rm{d}}^3 x}{(2 \pi)^{3/2}} {\rm e} ^{- i \vec{k} \cdot \vec{x}} \left[  \hat{E}_i \hat{E}_j+\hat{B}_i \hat{B}_j \right] := \hat{{\cal{S}}}_\lambda (\tau,\vec{k}),
    \label{eq:eom_Q}
\end{equation}
\noindent from the solution, we can evaluate the power spectrum for the gravitational waves
\begin{equation}
    \Big\langle  \hat{h}_\lambda (\tau, \vec{k}) \;\;\hat{h}_\lambda (\tau,\vec{k}^\prime)  \Big\rangle = \frac{4}{M_p^2 a^2} \Big\langle  \hat{Q}_\lambda (\tau, \vec{k})\;\; \hat{Q}_\lambda (\tau,\vec{k}^\prime)  \Big\rangle = \frac{2 \pi^2}{k^3} P_{\lambda} (k)\; \delta^{(3)}(\vec{k}+\vec{k}').
    \label{eq:PS}
\end{equation} 
From the power spectrum, we can extract the present fractional energy density. To do so, we need to account for the transfer functions from horizon re-entry to today. The modes in which we are interested become sub-Hubble again during radiation domination, and thus, we obtain the present fractional energy density  \cite{Caprini:2018mtu}
\begin{equation}
    \Omega_{\rm{GW}}(k) h^2  := \frac{h^2}{\rho_c} \frac{d \rho_{\rm{GW}}}{d \ln k} = \frac{\Omega_R h^2}{24} \sum_\lambda P_\lambda (k),
\end{equation}
where $\rho_c$ is the critical energy density at present time given by \( \rho_c=3 H_0^2 M_p^2 \), \(h\) parametrizes the uncertainty in the Hubble parameter in units of \(100\; {\rm{km}}\; {\rm{s}}^{-1}\;{\rm{Mpc}}^{-1}\), while $\Omega_R \simeq 4.15 \times 10^{-5}/ h^2$ is the fractional energy density in photons and neutrinos (as if they were relativistic) today.

\noindent The solution of eq.(\ref{eq:eom_Q}) is given by the sum of the vacuum contribution, which is the solution of the homogeneous equation, plus a sourced term. Therefore, we split  $\hat{Q}_\lambda(\tau,\vec{k})= \hat{Q}^{(v)}_\lambda(\tau,\vec{k}) + \hat{Q}^{(s)}_\lambda(\tau,\vec{k})$. These two contributions are statistically independent, hence, the final power spectrum will be given by a sum of the different contributions.

To compute the vacuum power spectrum, we expand the operator $\hat{Q}$ in its creation and annihilation components as
\begin{equation}
    \hat{Q}^{(v)}_\lambda (\tau,\vec{k})= q_\lambda (\tau,k) \; \hat{b}_\lambda (\vec{k})+q^*_\lambda (\tau,k) \; \hat{b}^{\dagger}_\lambda (\vec{k}),
\end{equation}
\noindent where $\hat{b}_\lambda^{\dagger} (\vec{k})$ creates a graviton of momentum $\vec{k}$ and helicity $2 \lambda$ out of $\ket{0}_{\rm{BD}}$. The equation of motion for the mode functions is
\begin{equation}
    \left( \frac{\partial^2}{\partial \tau^2}+k^2-\frac{a''}{a} \right) q_\lambda (\tau,\vec{k})=0,
\end{equation}
\noindent which are solved, imposing the initial Bunch-Davies condition, by
\begin{equation}
    q_\lambda(\tau, k)= \frac{{\rm{e}}^{- i k \tau}}{\sqrt{2k}} \left( 1- \frac{i}{k \tau} \right).
\end{equation}
Inserting this expression into eq.(\ref{eq:PS}) and taking the super-horizon limit, we obtain the vacuum power spectrum
\begin{equation}
    P_\lambda ^{(v)} (k)= \frac{H^2}{\pi^2 M_p^2},
\end{equation}
\noindent which is scale-invariant and unpolarized.
To obtain the sourced contribution, we express the solution of eq.(\ref{eq:eom_Q}) as
\begin{equation}
    \hat{Q}_\lambda^{(s)} (\tau, \vec{k})\; = \; \int^\tau {\rm{d}} \tau' G_k (\tau, \tau')\; \hat{{\cal{S}}}_\lambda (\tau', \vec{k}),
    \label{eq:sol_Q}
\end{equation}
\noindent where we have introduced the Green function $G_k (\tau,\tau')$ that satisfies
\begin{equation}
    \left( \frac{\partial^2}{\partial \tau^2}+k^2-\frac{a''}{a} \right) G_k (\tau,\tau')\;=\; \delta^{(3)}(\tau-\tau').
\end{equation}

\noindent Since we are interested in the retarded Green function, we can write it as
\begin{equation}
    G_k(\tau,\tau')= \tilde{G}_k (\tau,\tau')\, \, \Theta (\tau-\tau'),
\end{equation}
\noindent where $\Theta$ denotes the Heaviside step function.
To compute the power spectrum, we have to calculate the two-point function associated to the source operator $\hat{{\cal{S}}}_\lambda (\tau, \vec{k})$ defined in eq.(\ref{eq:eom_Q}). Substituting the expressions for the electric and magnetic field and taking the expectation value on the Bunch-Davies vacuum, we obtain
\begin{eqnarray}
        \langle {\cal{S}}_\lambda {(\tau',\vec{k})}  {\cal{S}}_\lambda {(\tau'',\vec{k}')} \rangle\;\;  =&& \frac{2 \delta^{(3)} (\vec{k}+\vec{k}')}{a(\tau') a(\tau'')M_p^2} \int \frac{{\rm{d}}^3 p}{(2 \pi)^3} \frac{1}{16} \sum_{\lambda_1,\lambda_2 }
  \left( 1+\lambda \lambda_1 \cos \theta \right)^2 \left( 1- \lambda \lambda_2 \frac{p \cos \theta - k}{\sqrt{k^2-2pk \cos \theta +p^2}} \right)^2  \nonumber \\
    &&  \left[ F_{E,\lambda_1} (\tau',p) F_{E,\lambda_2} (\tau',k-p) + F_{B,\lambda_1} (\tau',p) F_{B,\lambda_2} (\tau',k-p)\right]\nonumber \\
    && \left[  F _{E,\lambda_1} (\tau'',p) F _{E,\lambda_2} (\tau'',k-p) +  F _{B,\lambda_1} (\tau'',p) F _{B,\lambda_2} (\tau'',k-p)   \right]^*,
\end{eqnarray}
\noindent where $\theta$ is the angle between the integration momentum $p$ and the external momentum $k$. Furthermore, we have used the following property
\begin{equation}
   \left\vert \vec{\epsilon}_{\lambda_1}(\vec{p}_1) \cdot  \vec{\epsilon}_{\lambda_2}(\vec{p}_2) \right\vert^2 = \frac{1}{4} \left( 1- \lambda_1 \lambda_2 \frac{\vec{p}_1 \cdot \vec{p}_2}{|\vec{p}_1||\vec{p}_2|} \right)^2.
\end{equation}
Having the expression for the two-point function of the source operator, we can compute the sourced power spectrum using eq.(\ref{eq:PS})
\begin{eqnarray}
    P_\lambda^{(s)} (k)&&= \lim_{\tau \to 0} \frac{k^3}{4 \pi^2 a(\tau)^2 M_p^4} \int \frac{{\rm{d}}^3 p}{(2 \pi)^3} \sum_{\lambda_1,\lambda_2}
  \left( 1+\lambda \lambda_1 \cos \theta \right)^2 \left( 1- \lambda \lambda_2 \frac{p \cos \theta - k}{\sqrt{k^2-2pk \cos \theta +p^2}} \right)^2 \nonumber\\
  && \left\vert \int_{- \infty}^{\tau} d \tau' \frac{G_k (\tau,\tau')}{ a(\tau')} \Big[ A'_{\lambda_1} (\tau',p) A'_{\lambda_2} (\tau',\vert \vec{k}-\vec{p} \vert) + p \vert \vec{k}-\vec{p} \vert A_{\lambda_1} (\tau',p) A_{\lambda_2} (\tau',\vert \vec{k}-\vec{p} \vert) \Big] \right\vert^2.
\end{eqnarray}

\subsection{Retarded Green function}
\label{subsec:green_function}
In this section, we discuss the properties of the Green function in more detail following \cite{Garcia-Bellido:2023ser}. The retarded Green function $\tilde{G}_k(\tau,\tau')$ must satisfy the following conditions
\begin{equation}
    \left( \frac{\partial^2}{\partial \tau^2}+k^2-\frac{a''}{a} \right)\, \tilde{G}_k(\tau,\tau')=0,\;\;\;\tilde{G}_k(\tau,\tau)=0,\;\;\;\frac{d}{d \tau}\, \tilde{G}_k(\tau,\tau') \Big\vert_{\tau'=\tau}=1.
    \label{eq:equations_retarded_green}
\end{equation}
It is possible to construct a general solution satisfying this condition starting from two linearly independent solutions of the first of eq.(\ref{eq:equations_retarded_green}) 
\begin{equation}
    \left( \frac{\partial^2}{\partial \tau^2}+k^2-\frac{a''}{a} \right)\,F_i(\tau,k)=0,\;\;\;i=1,2.
    \label{eq:equations_F}
\end{equation}
For a given momentum $k$, deep inside the horizon, the mode behaves as in Minkowski spacetime, since the term $a''/a$ is negligible with respect to $k$. Therefore, we require that for asymptotically early times, the solutions are given by the usual flat-spacetime mode functions and we choose $F_1$ to approach the positive energy vacuum solution
\begin{equation}
    \lim_{\tau \to -\infty} F_1(\tau,k)=\, \frac{{\rm{e}}^{- i k \tau}}{\sqrt{2 k}}.
    \label{eq:initial_condition_F}
\end{equation}
If we now choose $F_2(\tau,k)=F_1^* (\tau,k)$, we have two linearly independent solutions. We can build the Wronskian and evaluate it when the asymptotic forms discussed above are valid, finding
\begin{equation}
    F_1'(\tau,k) F_2(\tau,k)-F_2'(\tau,k) F_1(\tau,k)=-\;i.
    \label{eq:Wronskian}
\end{equation}
It is straightforward to verify that the Wronskian is a constant of motion using eq.(\ref{eq:equations_F}). Therefore, eq.(\ref{eq:Wronskian}) is valid at all times. Using this condition, we can build a general solution for the retarded Green function
\begin{equation}
    \tilde{G}_k(\tau,\tau')\,=\, \frac{F_1(\tau,k)\, F_2(\tau',k)- F_2(\tau,k) \,F_1(\tau',k)}{F_1'(\tau',k)\, F_2(\tau',k)- F_2'(\tau',k) \, F_1(\tau',k)} = \frac{{\rm{Im}} \left[ F_1 (\tau,k) F_1^* (\tau',k) \right]}{{\rm{Im}} \left[ F_1' (\tau',k) F_1^* (\tau',k) \right]}.
    \label{eq:Green_function_solution}
\end{equation}
One can prove that this combination satisfies all the requirements of eq.(\ref{eq:equations_retarded_green}). This general solution is valid for arbitrary evolution of the scale factor, as long as we have a prolonged phase of inflation so that we can motivate the initial condition of eq.(\ref{eq:initial_condition_F}). In fact, one can show that it is possible to reproduce the known results for a de Sitter spacetime starting from (\ref{eq:Green_function_solution}).
Therefore, we can use this solution for our case, where the evolution of the scale factor is highly non-trivial during the inflationary transition. An important result that we will use for the numerical part concerns the late-time asymptotic behavior of the Green function. In the super-horizon limit given by $k \ll a(\tau) H(\tau)$, the most general solution is given by
\begin{equation}
    \lim_{\frac{k}{a(\tau) H(\tau)} \, \to\, 0} F(\tau,k)= a(\tau) \left[ c_1+c_2 \int_{\tau_*}^{\tau} \frac{{\rm{d}} \tau'}{a^2(\tau')} \right],
\end{equation}
where $c_1$ and $c_2$ are integration constants that generally depend on the specific past evolution of the scale factor and on $k$, while $\tau_*$ is a reference time after which the super-horizon approximation is valid for the given mode $k$. The combination of eq.(\ref{eq:Green_function_solution}) is however independent from these integration constants, therefore, the behavior of the retarded Green function in the super-horizon limit is independent from the details of the past evolution of the scale factor and is also $k-$independent. Thus, we have
\begin{eqnarray}
    \lim_{k\ll a(\tau) H(\tau),a(\tau') H(\tau')} \tilde{G}_k(\tau,\tau')&&= \lim_{k\ll a(\tau) H(\tau),a(\tau') H(\tau')} \frac{{\rm{Im}} \left[ F_1 (\tau,k) F_1^* (\tau',k) \right]}{{\rm{Im}} \left[ F_1' (\tau',k) F_1^* (\tau',k) \right]}\nonumber\\
    &&= a(\tau) a(\tau') \int_{\tau'}^\tau \frac{d \tau''}{a^2(\tau'')}.
\end{eqnarray}
We use this equation in the super-horizon regime to speed up the numerical integrations.

\section{The numerical setup}
\label{sec:numerical}

To solve the system of equations (\ref{eq:numeric_equations}) numerically, several methods have been used. One such approach consists of discretizing the backreaction integral \cite{Cheng:2015oqa,Notari:2016npn,DallAgata:2019yrr,Garcia-Bellido:2023ser}, where the integral is computed at each time step using the trapezoidal or parallelogram rule.

A second approach, applied in \cite{Sobol:2019xls,Gorbar:2021rlt}, uses the gradient expansion formalism. Then, there is the method introduced in \cite{Domcke:2020zez}, where the system of equations is solved iteratively, starting from an analytic approximation as the first iteration step.
Then, a full lattice computation was developed, which also includes the spatial gradients of the gauge field and the inflaton \cite{Caravano:2022epk,Caravano:2024xsb}. This lattice computation has been used to extract the scalar power spectrum.

In this work, we follow the first approach, building upon the numerical code presented in \cite{Garcia-Bellido:2023ser}.
We rescale the momenta to dimensionless variables that we use in the numerical code, the rescaling is defined in Appendix \ref{App:Code_variables}. We denote the dimensionless comoving momenta as $\tilde{k}$ and consider a set of $\tilde{k}_i$ for which we want to obtain the numerical solutions $A_\lambda (\tau,k_i)$ and over which we integrate to obtain the backreaction integral.
We take these momenta logarithmically equally spaced between $\tilde{k}_{\rm{min}}$ and $\tilde{k}_{\rm{max}}$, therefore, we have
\begin{equation}
    \tilde{k}_i=\tilde{k}_{\text{min}} \left( \frac{\tilde{k}_{\text{max}}}{\tilde{k}_{\text{min}}} \right)^{\frac{i-1}{i_{\text{max}}-1}},
    \label{eq:momenta_discretization}
\end{equation}
where we take $i_{\rm{max}} \simeq 500$ which allows for a good resolution of the GW power spectrum and for the backreaction integral. 
By setting the dispersion relation to zero in eq.(\ref{eq:A:xi}), we obtain the threshold between stability and instability (hence production) of a given mode. In code variables, this threshold is given by
\begin{equation}
    \tilde{k}_{\rm{thr}}(N) := c_F {\rm{e}}^{N-N_t} h(N) \frac{d \tilde{\theta}}{dN},
\end{equation}
where $N$ denotes the number of e-folds and $N_t$ the value of $N$ at which the first inflationary phase ends and the transition begins. 
We start the evolution of the equations about 10 e-folds before the beginning of the transition at $N_0$. For the lower bound, we require that all the $\tilde{k}_i$ are deep in the stable regime at the beginning of the run, while for the upper bound we take a value such that the production of modes for $\tilde{k}>\tilde{k}_{\rm{max}}$ is completely negligible. This translates into the following choices
\begin{equation}
    \tilde{k}_{\rm{min}}=2 \times 10^3 \; \tilde{k}_{\rm{thr}}(N_0),\;\;\; \tilde{k}_{\rm{max}}= 10^{-3} a(N_{\rm{end}})\; h(N_{\rm{end}}),
\end{equation}
where \( N_{\rm{end}} \) denotes the end of the last period of inflation. The integrand of the backreaction term $\langle \hat{\vec{E}} \cdot \hat{\vec{B}} \rangle$, expressed in code variables in eq.(\ref{eq:bacreaction_integrals_numeric}), generally requires regularization in the ultraviolet (UV) regime. Although such regularization might seem unnecessary at first glance, since the term \( \frac{d}{dN} \vert \tilde{A}_\lambda\vert^2 \) formally vanishes deep within the stable regime, small, non-zero numerical noise is always present. 
This noise is then amplified in the UV due to the factor $\tilde{k}^2$ in the integrand, combined with the fact that the comoving momentum spans many orders of magnitude in $k$-space. Consequently, it becomes essential to regularize the integral, which is typically achieved by imposing a hard momentum cutoff beyond which contributions are disregarded. It is possible to simply use a hard momentum cut-off because there is a clear separation of scales between a mode that is still in the vacuum state and a mode that is excited. Therefore, it is not necessary to employ more sophisticated renormalization procedures in curved spacetime.
The function $\tilde{k}_{\text{reg}}(N)$ defines the time-dependent momentum threshold beyond which gauge modes are excluded from the backreaction term.
Physically, at any given time, this threshold corresponds to the maximum comoving momentum that has ever entered the tachyonic regime at any earlier point in time, that is
\begin{equation}
    \tilde{k}_{\rm{reg}} (N) := {\rm{MAX}} \left[ \tilde{k}_{\rm{thr}}(N') \right],\;\;\; {\rm{for}}\;\; N' \in \left[N_0,N \right].
\end{equation}
The set of momenta considered here spans a range of about 15 orders of magnitude. It is therefore rather challenging to numerically evolve the equations for the mode functions of the gauge field and the auxiliary functions for the Green function. In fact, the equations contain terms given by \( \tilde{k}/(a h) \) that are many orders of magnitude larger than one for the largest modes in our set at the beginning of the evolution. These modes become dynamically relevant only when  \( \tilde{k}/(a h) \) has changed by many orders of magnitude from the initial value. Keeping track of these large hierarchies becomes very computationally expensive.
To address this problem, we proceed as in \cite{Garcia-Bellido:2023ser}. The strategy is to keep the modes frozen in the vacuum state until they come close to the instability threshold. To do so, we define the momentum scale below which the mode is kept fixed as $\tilde{k}_{\rm{vac}}$. Ideally, this scale should be as low as possible in order to keep the modes fixed for a longer time. But simultaneously, it should be much higher than the threshold scale, so that a given mode is safely in the vacuum regime when it begins to be free to evolve. We find a convenient value to be
\begin{equation}
    \tilde{k}_{\rm{vac}}(N):= 10^{3.5}\; \tilde{k}_{\rm{reg}}(N).
\end{equation}
The picture is as follows: a given mode $\tilde{k}_i$ is initially in its vacuum state and is kept frozen in the vacuum state until \( \tilde{k}_{\rm{vac}}(N) > \tilde{k}_i \). At this point, the mode is free to evolve according to its equation of motion, only after \( \tilde{k}_{\rm{reg}}(N)> \tilde{k}_i \) the mode is included in the backreaction integral and in the calculation of the GW power spectrum. The details of how this is done in practice are provided in Appendix \ref{app:cut-off}.

\section{Results}
\label{sec:results}
\begin{figure}[h]
\centering
\includegraphics[scale=1]{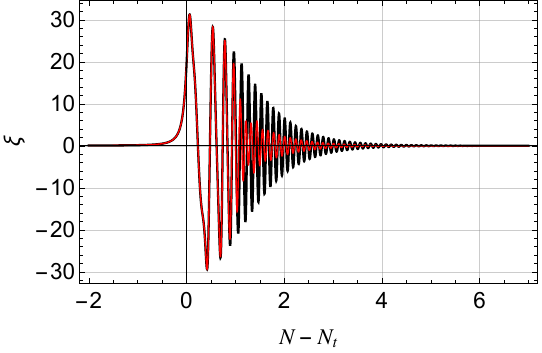}
\caption{Comparison between \(\xi\) in presence (red) and in absence (black) of the gauge field for $n_1 = \frac{70}{M_p^2} ,\, n_2 = \frac{500}{M_p^2} ,\, \gamma = 0.0067 ,\, r_\Lambda = 0.2,\; c_F=1.2,\; c_G=0$. As one can see, the backreaction slows the axions and dumps the oscillations. }
\label{fig:xi_comparison}
\end{figure}

In this section, we present the results obtained from the numerical scheme described above. Firstly, we find a non negligible backreaction on the axion field for this choice of parameters\footnote{Provided in the caption of fig. (\ref{fig:xi_comparison})} (we remind that only \( \theta\) is coupled to the gauge field) which slows the axion, damping its oscillations as visible in fig. (\ref{fig:xi_comparison}).

\begin{figure}[h]
\centering
\includegraphics[scale=1]{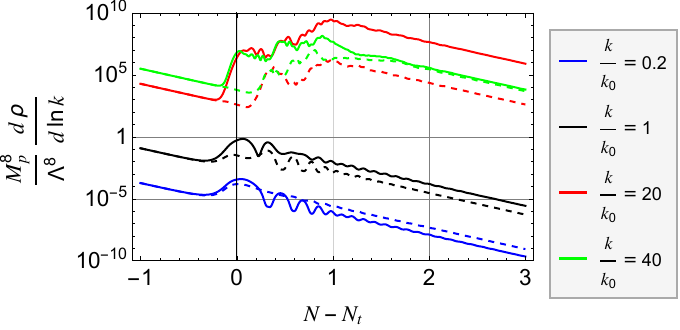}
\caption{Comparison between the spectral energy density for the gauge field as a function of \( (N-N_t) \) for different modes for left (solid) and right (dashed) polarizations. Except for the smallest wavelength modes showed in the figure, which does not contribute much to the GW power spectrum, the energy density for right-handed polarization is at least one order of magnitude smaller than the left-handed one asymptotically.
We remind that $k_0$ is the value of momentum that becomes super-horizon at the end of the first phase of inflation.
}
\label{fig:energy_densities_gauge}
\end{figure}

This behavior is well known (see, for example, \cite{Anber:2009ua}); it shortens the transient phase and, in turn, reduces the production of GWs. Figure (\ref{fig:energy_densities_gauge}) shows the spectral energy density of the gauge field for both polarizations. The total energy density, given by the sum of the two contributions, reads
\begin{eqnarray}
\label{eq:energy_density_gauge}
   \langle \hat{\rho} \rangle && = \left\langle \frac{\hat{E}_i \hat{E}_i + \hat{B}_i \hat{B}_i}{2} \right\rangle \nonumber \\ 
   && = \frac{1}{4 \pi^2 a^4}\int {\rm{d}} \ln k \; k^3 \sum_\lambda \left( |A'_\lambda(k)|^2 + k^2 |A_\lambda (k)|^2 \right)=\int {\rm{d}} \ln k \sum_\lambda \frac{d \rho}{d \ln k},
\end{eqnarray}

\begin{figure}
\centering
\includegraphics[scale=0.77]{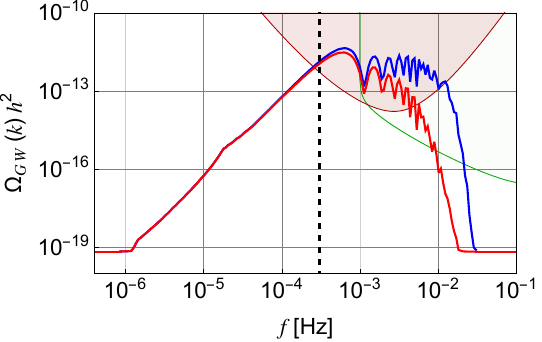}
\includegraphics[scale=0.76058]{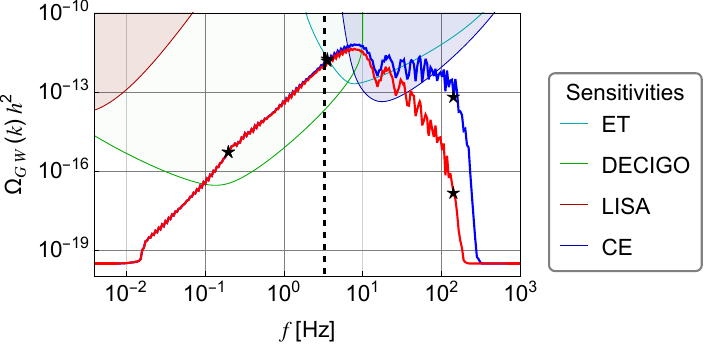}
\caption{Left panel: We show the GWs power spectrum obtained with the following parameters: $n_1=152.1/M_p^2,\;\; n_2=390/M_p^2,\;\; \gamma=0.0067\;\; r_\Lambda=0.51,\;\; c_F=1.6$. For these set of parameters, we have a second phase of inflation of about $27$ e-folds and therefore the peak of the power spectrum lies in the LISA range. The CMB phenomenology is given by $n_s\simeq 0.963,\;\;r\simeq 3 \times 10^{-3}$. \\
Right panel: GWs power spectrum for $n_1=70/M_p^2,\;\; n_2=500/M_p^2,\;\; \gamma=0.0067\;\; r_\Lambda=0.2,\;\;c_F=1.2$. In this case we have a second phase of inflation which lasts for about $17$ e-folds, therefore the peak lies at higher frequencies. The CMB phenomenology in this case is given by $n_s\simeq0.97,\;\;r\simeq 6\times10^{-4}$.
In both plots, the blue line represents the left-handed polarization and the red line the right-handed one. The black dashed line indicates $f_0$, the frequency corresponding to $k_0$, which is approximately $3 \times 10^{-4}\,{\rm Hz}$ in the left panel and $3.3\,{\rm Hz}$ in the right panel. In addition, we show the integrated sensitivity curves of future gravitational-wave detectors, reported in ~\cite{Schmitz:2020syl}.}
\label{fig:power_spectrum}
\end{figure}

\noindent where the expectation value is taken on the vacuum state and the expression written in terms of code variable is given in Appendix \ref{App:Code_variables}.
In the figure, one can see that the energy density of the gauge field is dominated by an intermediate range of modes, as the production already starts to decrease for the mode with $k = 40\,k_0$, compared to the one with $k = 20\,k_0$, where $k_0$ is the mode that becomes super-horizon at the end of the first phase of inflation.
As explained in Section \ref{sec:model}, contrary to the most studied cases, here we have a considerable production of both polarizations, however, for the intermediate modes, which experience the most efficient production, the right-handed polarization is at least one order of magnitude smaller than the left-handed one. It is therefore a valid approximation to neglect the contribution from the right-handed polarization in the computation of the GW power spectrum.
The exception to this is for the production of right-handed GW of wavenumber $k \gg k_0$, which are mostly produced by right-handed gauge modes of comparable large $k$, as is explained further in the rest of this section.
Finally, in fig. (\ref{fig:power_spectrum}), we show the computed GW power spectra for two different sets of parameters, plotting \(\Omega_{\rm{GW}} h^2\) as a function of frequency. Qualitatively, we see the same behavior of the power spectra. We have an approximately cubic growth before reaching the maximum at $f\gtrsim f_0$, where $f_0$ is the frequency that becomes super-horizon at $N_t$. Therefore, we find a very distinctive imprint of the transition in the GW power spectrum, with the peak pointing directly at the transition scale. 

We note that the shape of the gravitational wave power spectrum is reminiscent of the evolution of the parameter $\xi$ during the transition. Specifically, the spectrum exhibits a peak at the transition scale, followed by damped oscillations. 
This behavior can be understood through a series of considerations, as discussed in \cite{Garcia-Bellido:2023ser}. As detailed in Appendix~\ref{app:loop_integral}, the production of gravitational waves with momentum $k$ is governed by a \textit{loop contribution} (Eq.~\ref{eq:loop_integral}), in which two gauge modes with momenta $\vec{p}$ and $\vec{q}$ combine to produce a gravitational wave, subject to the momentum conservation condition $\vec{p} + \vec{q} = \vec{k}$.
When $\xi$ is slowly increasing, as before the transition, a gravitational wave mode with momentum $k$ that exits the horizon at $N_k$ is predominantly sourced by gauge modes crossing the horizon at the same time. Since $\xi$ increases with time in this regime, $\Omega_{\mathrm{GW}}(k)$ also increases with $k$. This behavior changes once $\xi$ becomes approximately constant or slightly decreases, as occurs after the transition. Consequently, the overall evolution of $\xi$ naturally leads to a peak in the power spectrum around the frequency $f_0$.

Another feature of the power spectra that we find is the violation of parity, typical of ALPs coupled to a gauge field. In this case, we have a non chiral power spectrum for $f\lesssim f_0$, while after the transition scale, the right-handed GWs decay at a faster rate with respect to the other polarization. To analyze the chiral behavior of the power spectrum, we define the chirality parameter at a certain frequency $f$ as 
\begin{equation}
    \Delta \chi (f) := \frac{P_+ (f)\, -\, P_- (f)}{P_+ (f) \,+\, P_- (f)},
    \label{eq:Chirality}
\end{equation}
and for the three frequencies marked in the right panel of fig. (\ref{fig:power_spectrum}), we find
$\Delta \chi (f_1) \simeq 0.007,\; \Delta \chi (f_2) \simeq 0.13,\; \Delta \chi (f_3) \simeq 0.999$, with $f_1 \simeq 0.2\; {\rm{Hz}},\; f_2 \simeq 3.6 \; {\rm{Hz}}, \; f_3 \simeq 142 \; {\rm{Hz}}$.
The pattern of parity violation can be understood by examining the loop contributions that source the gravitational waves \cite{Garcia-Bellido:2023ser}. 
Gravitational waves with $\vert \vec{k} \vert \lesssim k_0$ are produced during the phase in which $\xi$ is increasing. Their generation is mainly sourced by gauge modes with $\vert \vec{p} \vert,\, \vert \vec{q} \vert \gtrsim k_0$, since these modes are amplified later than those of order $k_0$, when $\xi$ is larger and the production more efficient. In this regime, momentum conservation requires the two gauge field momenta to be nearly anti-aligned. As a result, the corresponding state carries almost no net helicity, leading to an approximately equal production of the two GW polarizations.
On the other hand, gravitational waves with $\vert \vec{k} \vert \gtrsim k_0$ are predominantly sourced by gauge modes with $\vert \vec{p} \vert,\, \vert \vec{q} \vert \lesssim k_0$, as they are produced during the stage in which $\xi$ undergoes damped oscillations. In this regime, the initial helicity is no longer nearly vanishing, leading to the parity violation observed in fig. (\ref{fig:power_spectrum}). 
In this case, left-handed gauge modes predominantly source left-handed gravitational waves, and analogously for the right-handed ones. 
Since right-handed gauge modes are produced less efficiently, the right-handed gravitational wave component is suppressed. 
Including the right-handed polarization of the gauge field in the calculation would thus slightly enhance the right-handed power spectrum, resulting in a mild reduction of $\Delta \chi$ for $f \gg f_0$. 
A more detailed discussion of these effects is presented in Appendix~\ref{app:loop_integral}.
It is important to remind that these results are obtained under the assumption of homogeneous axion fields, with the backreaction treated accordingly. Verifying this requires a study of the scalar perturbations, which is left for future work.

\section{Conclusions}
\label{sec:conclusion}
Most of the existing literature on aligned natural inflation has focused on the compatibility of the model with quantum gravity embeddings. In this work, we study instead the phenomenological predictions of the model when coupled to a U(1) gauge field.
Usually, the heavy direction is integrated out, and inflation is assumed to take place along the trajectory which terminates on the minimum of the potential. This class of solutions resembles the single-field natural inflation model, and they are ruled out by the CMB phenomenology \cite{Planck:2018jri}. In \cite{Peloso:2015dsa}, it was noted that the model also possesses a second class of inflationary trajectories which are not connected to the minimum. These trajectories, that we define \textit{metastable}, stream off from a saddle point of the potential and inflation terminates because of an instability in the direction orthogonal to the inflationary path. The metastable solutions present a suppressed tensor-to-scalar ratio, since inflation takes place around a saddle point where the potential is flatter. In \cite{Peloso:2015dsa}, this class of solutions has been studied numerically, contrary to \cite{Greco:2024ngr}, where an analytic study has been performed, finding approximate expressions for the scalar tilt and the tensor-to-scalar ratio of the vacuum scalar perturbations. Moreover, the presence of a transition between the second and the first class of inflationary trajectories has been pointed out, but it was not clear if this possibility was a peculiar feature of the choice of parameters or a general prediction of the model.

In this work, we study in more detail the second phase of inflation and its occurrence. 
Although we are not able to describe analytically the phase of transition between the two inflationary regimes, we derive a simple analytic result which approximates the number of e-folds obtained in the second phase of inflation for a given set of parameters, given by eqs. (\ref{eq:e_folds_analytic}), (\ref{eq:phi0}). From this, we are able to show (see fig. (\ref{fig:e-folds-extra})) that the presence of a composite inflationary solution which takes place along the two different classes of trajectories is a generic property of the model and not a peculiar feature which takes place only in a small fraction of the parameter space. 

Having established the generic existence of transitions between inflationary trajectories, we now examine a possible phenomenological signature associated with them. To do so, we consider the original aligned model coupled to a U(1) gauge field. In fact, the presence of a topological Chern-Simons coupling term between axions and gauge fields is usually ubiquitous in every axion realization (see for instance \cite{DiLuzio:2020wdo}).
To simplify the treatment, we consider only the coupling of the gauge field with $\theta$. As discussed in more detail in Section~\ref{subsec:equations-of-motion}, the coupling of the axions to the U(1) field is expected to be proportional to that of the confining sector responsible for generating the axion masses. This implies a suppressed coupling of the U(1) field to the light eigenstate. In presence of alignment, the heavy mode is almost coincident with $\theta$ (see eqs. (\ref{eq:heavy_eigenvec}),(\ref{eq:light_eigenvec})) leading to the above simplification.

We numerically evolve the coupled equations of the axions and the gauge field mode functions for a set of comoving momenta, including the backreaction of the gauge field on $\theta$. From this, we can compute the GW power spectra shown in fig. (\ref{fig:power_spectrum}). As a  result, we find a strong enhancement with respect to the vacuum value and a very distinctive shape of the power spectra. Indeed, we have a strongly scale-dependent power spectrum, peaking at the scale of the inflationary transition and then decaying in an oscillating behavior and possibly detectable from future GWs experiments as shown in fig. (\ref{fig:power_spectrum}). The decreasing part of the power spectrum is strongly chiral, since the right-handed polarization falls off more rapidly than the left-handed one.
It is important to stress that the specific choice of parameters adopted here serves primarily as an illustration. The central point is that the model is capable of producing this characteristic GW power spectrum, with the position of the peak spanning a broad range of frequencies depending on the duration of the second phase of inflation. 
Another well-studied mechanism that can generate a strongly scale-dependent tensor signal is provided by the so-called spectator axion models \cite{Namba:2015gja}. In this scenario, an axion field coupled to a gauge field evolves on an inflationary background driven by the inflaton. By tuning the axion decay constant relative to the Hubble rate, the axion can be made to roll only for a limited number of e-folds (typically $\sim 2$–$5$). As a result, the spectrum of gravitational waves is strongly sensitive to the modes that cross the horizon during the rolling phase of the axion.
A key feature of this class of models is that such behavior requires a delicate tuning of the axion mass with respect to the Hubble rate. 
In general, any model that produces a peak in the GW spectrum necessarily involves a characteristic scale, since the position of the peak defines it. In the case of aligned natural inflation, this scale is not introduced explicitly but instead emerges non-trivially from the model parameters, which determine the duration of the second inflationary stage.
This distinction underscores the robustness of the predictions associated with aligned natural inflation, as the occurrence of a second inflationary phase is a generic feature of the model. Although there may exist solutions in which this phase is so short as to become negligible, the framework naturally accommodates the generation of phenomenologically relevant GW signatures.

In this work, we have assumed homogeneity in the axion fields. Although the code correctly accounts for the backreaction of the gauge field, the equations have been solved under the approximation of a homogeneous backreaction. A complete analysis would require computing scalar perturbations to test the validity of this assumption. We plan to pursue this direction in future work, laying the basis for the phenomenology of inflationary transitions in aligned natural inflation in the current manuscript.

\appendix
\section{Code variables}
\label{App:Code_variables}
In this section, we write the rescaled variables that we use in the numerical integration of the equations. Firstly, as a time variable we use the number of e-folds, defined as
\begin{equation}
    N := \int_{t_0}^t {\rm{d}}t' H(t'),
\end{equation}
where $t_0$ is some reference initial time that we take when the CMB modes are deep inside the horizon. We normalize the scale factor so that it is equal to $1$ at $N=N_t$ which marks the end of the first phase of inflation, so we have
\begin{equation}
    a(N)= {\rm{e}}^{N-N_t}, \;\,\; \tau= \tau_0 + \int_{N_0}^N \frac{{\rm{d}}N'}{H(N') {\rm{e}}^{N'-N_t}}.
\end{equation}
Then, we rescale the dimensional variables to dimensionless ones
\begin{eqnarray}
   & \Tilde{\theta} := \sqrt{n_2} \theta, \;\; \Tilde{\rho} := \sqrt{n_2} \rho, \;\; \tilde{V} := \frac{1+r_\Lambda}{\Lambda^4} V,\nonumber\\
   & h := \left( \frac{\Lambda^2}{M_p \sqrt{3 (1+r_\Lambda)}} \right)^{-1} H, \;\; \tilde{k} := \left( \frac{\Lambda^2}{M_p \sqrt{3 (1+r_\Lambda)}} \right)^{-1} k,\;\;\tilde{A}_\lambda := \sqrt{2k} e^{ik(\tau -\tau_0)} A_\lambda,
   \label{eq:dimensionless_rescalings}
\end{eqnarray}
we have also rescaled the mode functions $A_\lambda(\tau,k)$ to eliminate the fast oscillating behavior inside the horizon. Now, we can rewrite the equations of motion in terms of dimensionless variables and using the number of e-folds as a time variable, obtaining
\begin{equation}
\left\{
\begin{aligned}
&\frac{d^2 \tilde{\theta}}{d N^2} + \left\{ 1 + \frac{2 \tilde{V}}{h^2} - \frac{1}{6 \tilde{n}_2} \left[ \left( \frac{d \tilde{\theta}}{dN} \right)^2 + \left( \frac{d \tilde{\rho}}{dN} \right)^2 \right] \right\} \frac{d \tilde{\theta}}{d N} + \frac{3 \tilde{n}_2}{h^2} \partial_{\tilde{\theta}} \tilde{V} + J_\theta = 0, \\
&\frac{d^2 \tilde{\rho}}{d N^2} + \left\{ 1 + \frac{2 \tilde{V}}{h^2} - \frac{1}{6 \tilde{n}_2} \left[ \left( \frac{d \tilde{\theta}}{dN} \right)^2 + \left( \frac{d \tilde{\rho}}{dN} \right)^2 \right] \right\} \frac{d \tilde{\rho}}{d N} + \frac{3 \tilde{n}_2}{h^2} \partial_{\tilde{\rho}} \tilde{V} + J_\rho = 0, \\
&\frac{d h}{dN} + \frac{h}{6 \tilde{n}_2} \left[ \left( \frac{d \tilde{\theta}}{dN} \right)^2 + \left( \frac{d \tilde{\rho}}{dN} \right)^2 \right] - \frac{2 \tilde{V}}{h} + 2h = 0, \\
&\frac{d^2 \tilde{A}_\lambda}{d N^2} + \left\{ \frac{2 \tilde{V}}{h^2} - \frac{1}{6 \tilde{n}_2} \left[ \left( \frac{d \tilde{\theta}}{dN} \right)^2 + \left( \frac{d \tilde{\rho}}{dN} \right)^2 \right] -1 - 2i \frac{\tilde{k}}{a h} \right\} \frac{d \tilde{A}_\lambda}{d N} - \lambda \frac{\tilde{k}}{a h} \left( c_F \frac{d \tilde{\theta}}{dN} + c_G \frac{d \tilde{\rho}}{dN} \right) \tilde{A}_\lambda = 0,
\end{aligned}
\right.
\label{eq:numeric_equations}
\end{equation}
where we have substituted $\vec{E} \cdot \vec{B}$ in eqs.(\ref{eq:theta}), (\ref{eq:rho}) with its vacuum expectation value given by
\begin{eqnarray}
   && J_{\theta} := - \sqrt{n_2} \frac{\langle \vec{E} \cdot \vec{B} \rangle}{F H^2}  = \frac{\Lambda^4}{M_p^4}  \frac{c_F \tilde{n}_2}{24 \pi^2 (1+r_\Lambda) a^3 h} \int {\rm{d}} \ln k \; \tilde{k}^3 \frac{\partial}{\partial N} \left[ |\tilde{A}_+|^2 - |\tilde{A}_-|^2  \right],\nonumber\\
    && J_{\rho} := - \sqrt{n_2} \frac{\langle \vec{E} \cdot \vec{B} \rangle}{G H^2}  = \frac{\Lambda^4}{M_p^4}  \frac{c_G \tilde{n}_2}{24 \pi^2 (1+r_\Lambda) a^3 h} \int {\rm{d}} \ln k \; \tilde{k}^3 \frac{\partial}{\partial N} \left[ |\tilde{A}_+|^2 - |\tilde{A}_-|^2  \right].
    \label{eq:bacreaction_integrals_numeric}
\end{eqnarray}
We also report here the energy density of the gauge fields of eq.(\ref{eq:energy_density_gauge}) written in code variables
\begin{equation}
\rho_g= \left(\frac{\Lambda^4}{M_p^2}\right)^2 \frac{1}{18 (2 \pi)^2 (1+r_\Lambda)^2} \int {\rm{d}} \ln k \; \frac{\tilde{k}^4}{a^4} \sum_\lambda \left( \frac{a^2 h^2}{\tilde{k}^2} \left| \frac{d\tilde{A}_\lambda}{d N} -\frac{i \tilde{k}}{a h} \tilde{A}_\lambda \right|^2+|\tilde{A}_\lambda|^2 \right).
\end{equation}
We numerically integrate the system of equations (\ref{eq:numeric_equations}). At each time step, we compute the backreaction integral $J_{\theta}$ ($J_{\rho}$ vanishes since we are considering $c_G=0$) over the discrete set of momenta, as discussed in Section \ref{sec:numerical}, by using the trapezoid rule. We use as integration variable $\ln \tilde{k}$ so that the integration step is constant for all points. Therefore, we have
\begin{equation}
\int_{\tilde{k}_{\text{min}}}^{\tilde{k}_{\text{max}}} {\rm{d}} \tilde{k}\; \mathcal{I}(\tilde{k}) 
\Rightarrow 
\frac{\Delta \ln \tilde{k}}{2} 
\left(
\tilde{k}_{\text{min}} \mathcal{I}_{\tilde{k}_{\text{min}}} 
+ \tilde{k}_{\text{max}} \mathcal{I}_{\tilde{k}_{\text{max}}} 
+ 2 \sum_{i=1}^{i_{\text{max}}-1} \tilde{k}_i \mathcal{I}_{\tilde{k}_i}
\right)
\end{equation}
The boundary terms are further disregarded to speed up the calculation, since the integrand assumes negligible values there. We start the numerical evolution at $N=N_0$ which is taken to be about 10 e-folds before the end of the first phase of inflation. At $N=N_0$, all the modes are still deep inside the horizon, therefore we take Bunch-Davies initial conditions for the mode functions. For the axion fields and the Hubble rate, we take their value at $N_0$ given by the background evolution. This is justified by the fact that we do not have gauge quanta production before the transition, since $\xi$ is effectively zero, therefore we can only consider the background evolution. Therefore, for the mode functions initials conditions, we have
\begin{equation}
    \tilde{A}_\lambda (N_0,k)=1,\;\;\; \frac{d \tilde{A}_\lambda (N_0,k)}{dN} \Big\vert_{N=N_0}=0.
\end{equation}

Now, we turn our attention to the computation of the retarded Green function that we need to numerically evaluate the GW power spectrum. Written in terms of $N$ as a time varible, eq.(\ref{eq:eom_Q}) becomes
\begin{equation}
    \hat{O}_N \hat{Q}_\lambda (N,\vec{k})= \frac{\hat{S}_\lambda (N,\vec{k})}{H^2 {\rm{e}}^{2(N-N_0)}}, 
\end{equation}
where we have defined the differential operator $\hat{O}_N$ as
\begin{eqnarray}
{\hat O}_N :=   \frac{d^2}{d N^2} + \left( 1 + \frac{1}{H} \frac{d H}{d N} \right) \frac{d}{d N} + \left[ \frac{k^2}{H^2 {\rm e}^{2 \left( N - N_t \right)}} - 2 - \frac{1}{H} \frac{d H}{d N} \right].
\end{eqnarray}
In terms of the Green function, we have
\begin{equation}
{\hat O}_N {\cal G}_k \left( N ,\, N' \right) = \delta \left( N - N' \right) \;\;,\;\; 
{\cal G}_k \left( N ,\, N' \right) := \tilde{\cal G}_k \left( N ,\, N' \right) \theta \left( N - N' \right),
\end{equation} 
which brings us to the formal solution for the canonical operator $\hat{Q}_\lambda(N,\vec{k})$
\begin{equation}
\hat{Q}_\lambda ( N ,\, \vec{k} ) = \int^N {\rm{d}} N' \tilde{\cal G}_k \left( N ,\, N' \right)
\frac{\hat{S}_\lambda \left( N' ,\, \vec{k} \right)}{H^2 ( N ') \, {\rm e}^{2 \left( N' - N_t \right)} } .
\end{equation} 
This solution coincides with (\ref{eq:sol_Q}), thus, we can immediately relate the two expressions for the Green function written in terms of $N$ and $\tau$
\begin{equation}
    \tilde{{\cal G}}_k (N,N')=H(N') {\rm e}^{N'-N_t} \tilde{G}_k \left(\tau(N),\tau'(N')\right).
\end{equation}
Now, we can write the expression for the power spectrum in terms of $N$ and $\tilde{{\cal G}}_k$ as
\begin{eqnarray} 
&& \!\!\!\!\!\!\!\!  \!\!\!\!\!\!\!\!  \!\!\!\!\!\!\!\! \!\!\!\!\!\!\!\!  \!\!\!\!\!\!\!\! 
P_\lambda^{(s)} \left( k \right) =  \frac{k^3 {\rm e}^{-2 \left( N-N_t\right)}}{16 \pi^4 M_p^4} \sum_{\lambda_1,\lambda_2} \int_0^\infty {\rm{d}} p \, p^2 \int_{-1}^1 {\rm{d}} \cos \theta   \left( 1 + \lambda \lambda_1 \cos \theta \right)^2 \left( 1 - \lambda \lambda_2 \frac{p \, \cos \theta - k}{\sqrt{k^2-2 p k \cos \theta + p^2}} \right)^2  \nonumber\\
&&  \!\!\!\!\!\!\!\!  \!\!\!\!\!\!\!\! \!\!\!\!\!\!\!\!  \!\!\!\!\!\!\!\! 
\left\vert \int_0^N {\rm{d}} N'  \, \frac{\tilde{\cal G}_k \left( N ,\, N' \right)}{ {\rm e}^{N'-N_t}} \, \left[ \frac{d A_{\lambda_1} \left( N' ,\, p \right)}{d N'} \frac{d A_{\lambda_2} ( N' ,\, \vert \vec{k} - \vec{p} \vert )}{d N'} + \frac{p \,\vert \vec{k} - \vec{p} \vert}{H (N')^2 {\rm e}^{2(N'-N_t)}}  A_{\lambda_1} \left( N' ,\, p \right) A_{\lambda_2} ( N' ,\, \vert \vec{k} - \vec{p} \vert ) \right] \right\vert^2 \nonumber\\ 
\end{eqnarray}
where we have performed the trivial angular integration given by the rotation of $\vec{p}$ at fixed $\theta$. At this point, we perform a change of variables, defining
\begin{equation}
    p := k (X+Y),\;\;\; \vert \vec{k}-\vec{p} \vert := k (X-Y),\;\;\; \Rightarrow \cos \theta = \frac{1+ 4 XY}{2(X+Y)}.
\end{equation}
In terms of these new variables, the integral becomes


\begin{eqnarray}
    && \!\!\!\!\!\!\!\!  \!\!\!\!\!\!\!\!  \!\!\!\!\!\!\!\! \!\!\!\!\!\!\!\!  \!\!\!\!\!\!\!\! 
    P_\lambda ^{(s)} (k) = \frac{k^6 {\rm{e}}^{-2(N-N_t)}}{128 \pi^4 M_p^4} \sum_{\lambda_1,\lambda_2}\int_{1/2}^{\infty} {\rm{d}} X\; \int_{-1/2}^{1/2} {\rm{d}}Y \frac{{\cal{P}}_{\lambda \lambda_1 \lambda_2}  }{X^2-Y^2} \nonumber\\
    &&  \!\!\!\!\!\!\!\!  \!\!\!\!\!\!\!\! \!\!\!\!\!\!\!\!  \!\!\!\!\!\!\!\! 
\left\vert \int_0^N {\rm{d}} N'  \, \frac{\tilde{\cal G}_k \left( N ,\, N' \right)}{ {\rm e}^{N'-N_t}} \, \left[ \frac{d A_{\lambda_1} \left( N' ,\, k (X+Y) \right)}{d N'} \frac{d A_{\lambda_2} ( N' ,\, k (X-Y) )}{d N'} + \frac{k^2 (X^2-Y^2)}{H (N')^2 {\rm e}^{2(N'-N_t)}}  A_{\lambda_1} \left( N' ,\,k (X+Y) \right) A_{\lambda_2} \left( N' ,\, k (X-Y) \right) \right] \right\vert^2, \nonumber\\
\end{eqnarray}
where ${\cal{P}}_{\lambda \lambda_1 \lambda_2}$ is the polarization factor given by

\begin{equation}
\left\{
\begin{aligned}
& {\cal{P}}_{+++}= \left( 1- 4 Y^2  \right)^2 \left( 1+2X  \right)^4= {\cal{P}}_{---} , \\
& {\cal{P}}_{++-}= \left( 1- 4 X^2  \right)^2 \left( 1+2 Y  \right)^4= {\cal{P}}_{--+}, \\
& {\cal{P}}_{+-+}= \left( 1- 4 X^2  \right)^2 \left( 1-2 Y  \right)^4= {\cal{P}}_{-+-}, \\
& {\cal{P}}_{+--}=  \left( 1-4Y^2  \right)^2 \left( 1- 2 X  \right)^4  = {\cal{P}}_{-++},
\end{aligned}
\right.
\label{eq:polarization_factor}
\end{equation}

This integral can be visualized as a loop integral with internal momenta given by $\vec{p}$ and $\vec{q}= \vec{k}-\vec{p}$.
Now, we rewrite the integral in terms of the dimensionless quantities defined in \ref{eq:dimensionless_rescalings}, obtaining
\begin{eqnarray}
&& \!\!\!\!\!\!\!\!  \!\!\!\!\!\!\!\!  \!\!\!\!\!\!\!\! \!\!\!\!\!\!\!\!  \!\!\!\!\!\!\!\! 
P_\lambda^{(s)} \left( k \right) =  \frac{\Lambda^8}{4608 (1+r_\Lambda)^2 \pi^4 M_p^8 a^2(N)} \sum_{\lambda_1,\lambda_2}\;\int_{1/2}^\infty {\rm{d}} X  \int_{-1/2}^{1/2} {\rm{d}} Y  \frac{{\cal{P}}_{\lambda \lambda_1 \lambda_2}}{(X^2-Y^2)^2} \nonumber\\
&&  \Bigg| \int_0^N {\rm{d}} N'  \, \frac{\tilde{k}^2 \tilde{\cal G}_k \left( N ,\, N' \right)}{ {\rm e}^{N'-N_t}} \,  {\rm{e}}^{- 2 i \tilde k X \int^{N'}_0 \frac{d n}{h(n)\, a(n)}}  \Bigg\{ \frac{d \tilde{A}_{\lambda_1} (N', \tilde{k}(X+Y))}{d N'} \, \frac{d \tilde{A}_{\lambda_2} (N', \tilde{k} (X-Y))}{d N'} \nonumber\\
&& - \frac{i \tilde{k}}{ a(N') \, h(N')} \Bigg[ (X+Y) \, \tilde{A}_{\lambda_1}(N', \tilde{k} (X+Y)) \, \frac{d \tilde{A}_{\lambda_2}(N', \tilde{k} (X-Y))}{d N'}\nonumber\\  
&&+ (X-Y) \, \tilde{A}_{\lambda_2}(N', \tilde{k} (X-Y)) \, \frac{d \tilde{A}_{\lambda_1} (N', \tilde{k} (X+Y))}{d N'} \Bigg]  \Bigg\} \Bigg|^2 \nonumber \\
&& := \int_{1/2}^{\infty} {\rm{d}} X \int_{-1/2}^{1/2} {\rm{d}} Y {\cal C} (X,Y).
\label{eq:loop_integral}
\end{eqnarray}
Now, we have to write the equation in code variable for the Green function. We have to follow the steps outlined in subsection \ref{subsec:green_function}, therefore we find the solution of the homogenous equation $\hat{O}_N {\cal{F}} (N,k)=0$. As we have done for the gauge field mode functions, we rescale ${\cal{F}}$ to eliminate the initial fast oscillating behavior, and we impose initial condition \ref{eq:initial_condition_F}. Hence, we define
\begin{equation}
    \tilde{{\cal{F}}} := \sqrt{2k} e^{ik(\tau -\tau_0)} {\cal{F}},\;\;\tilde{\cal{F}}(N_0,\tau) =1,\;\;\frac{d \tilde{\cal{F}}(N,\tau)}{dN} \Big|_{N_0}=0.
\end{equation}
$\tilde{\cal{F}}$ satisfies the equation of motion
\begin{equation}
    \frac{d^2 \tilde{{\cal{F}}}}{dN^2} + \left\{-1 - \frac{1}{6 \tilde{n}_2} \left[ \left( \frac{d \tilde{\theta}}{dN} \right)^2 + \left( \frac{d \tilde{\rho}}{dN} \right)^2 \right] +\frac{2 \tilde{V}}{h^2} - \frac{2 i \tilde{k}}{a h}  \right\} \frac{d \tilde{\cal{F}}}{dN} + \left\{ \frac{1}{6 \tilde{n}_2}  \left[ \left( \frac{d \tilde{\theta}}{dN} \right)^2 + \left( \frac{d \tilde{\rho}}{dN} \right)^2 \right]-\frac{2 V_\text{num}}{h^2}  \right\} \tilde{\cal{F}}=0.
    \label{eq:eom_F_tilde}
\end{equation}
From $\tilde{\cal{F}}$ we extract the Green function as
\begin{eqnarray}
   \tilde{\cal{G}}_k(N,N') && = \frac{{\rm{Im}} [{\cal{F}}(N,k) {\cal{F}}^*(N',k)]}{{\rm{Im}} \left[ \frac{d {\cal{F}}(N',k)}{dN'} {\cal{F}}^*(N',k) \right]}\nonumber\\
   &&= \frac{\text{Im} \left[  \text{e}^{- i \tilde{k}\int_{N'}^N \frac{dn}{h(n) a(n)}} \tilde{F}(N,k) \tilde{F}^* (N',k) \right]}{\text{Im} \left[ \frac{d \tilde{F}(N',k)}{dN'} \tilde{F}^*(N',k) \right]-\frac{\tilde{k}}{h(N') a(N')} |\tilde{F}(N',k)|^2}.
\end{eqnarray}
As derived previously, in the super-horizon limit, the Green function behaves as
\begin{equation}
    \lim_{\tilde{k} \ll a(N) h(n), a(N') h(N')}\tilde{\cal{G}}_k(N,N')=h(N') a^2(N') a(N) \int_{N'}^N \frac{{\rm{d}} N''}{h(N'') a^3(N'')}.
\end{equation}
Indeed, we use this asymptotic behavior in the numerical evaluation of the power spectrum to speed-up the calculations as highlighted in Subsection \ref{subsec:green_function}.

\section{Numerical implementation of the cut-off}
\label{app:cut-off}
In this appendix, we explain the implementation of the cut-off in momentum space used for the regularization of the backreaction integral and for the mode functions of the gauge field and the auxiliary functions ${\tilde{\cal{F}}}$.
For what concerns the backreaction integral, we build a regularization function $R (N, \tilde{k})$ in order to regularize the integral as
\begin{equation}
    {\cal{I}}_{\rm{reg}} (\tilde{k}) := R (N, \tilde{k}) {\cal{I}} (\tilde{k}).
\end{equation}
For the choice of this function, we follow \cite{Garcia-Bellido:2023ser}. $R$ should behave as a Heaviside step function, but the numerical differential equation solver in \textit{Mathematica} works better with analytic functions. Therefore, we define
\begin{equation}
    R(N,\tilde{k}) := \frac{1}{2} \left\{  \tanh\left[10^3 \ln \left(\frac{\tilde{k}_{\rm{reg}}}{\tilde{k}} \right) \right] +1 \right\}.
\end{equation}
This function is effectively indistinguishable from a step function. The function $\tilde{k}_{\rm{reg}}(N)$ has to be solved along with the equations of motion. Initially, $\tilde{k}_{\rm{thr}}(N)$ evolves monotonically with $N$ and therefore we have simply
\begin{equation}
    \tilde{k}_{\rm{reg}}(N)= \tilde{k}_{\rm{thr}}(N),
\end{equation}
when we reach the transition, $\tilde{k}_{\rm{thr}}(N)$ starts to oscillate, however, $\tilde{k}_{\rm{reg}}(N)$ has to remain a monotonous increasing function of $N$ being the greater value of $\tilde{k}_{\rm{thr}}(N)$ ever reached before that moment.
To implement this, we define a list $\tilde{\mathcal{K}}$ in \textit{Mathematica}. When $\tilde{k}_{\text{thr}}(N)$ changes monotonicity, we program \textit{Mathematica} to save the absolute value of the threshold at this time and we add it to $\tilde{\mathcal{K}}$. We define a discrete variable $\tilde{k}_{\text{loc,max}}(N)$ that is initialized as the value of $\tilde{k}_{\text{thr}}(N_0)$ but that does not obey to any differential equation. At every moment in time, this variable is equal to the maximum value of $\tilde{\mathcal{K}}$. Thus, we define $\tilde{k}_{\text{reg}}$ as
\begin{eqnarray}
    \tilde{k}_{\text{reg}}(N) &&= \Bigg\vert \frac{1}{2} \left\{   \tanh \left[ 10^3 \ln \left( \frac{|\tilde{k}_{\text{thr}}(N)|}{\tilde{k}_{\text{loc,max}}(N)} \right)  \right] +1 \right\} \tilde{k}_{\text{thr}}(N)\nonumber \\
    && + \frac{1}{2} \left\{   \tanh \left[ -10^3 \ln \left( \frac{|\tilde{k}_{\text{thr}}(N)|}{\tilde{k}_{\text{loc,max}}(N)} \right)  \right] +1 \right\} \tilde{k}_{\text{loc,max}}(N)     \Bigg\vert.
    \label{eq:k_regulator}
\end{eqnarray}

\begin{figure}
\centering
\includegraphics[scale=1]{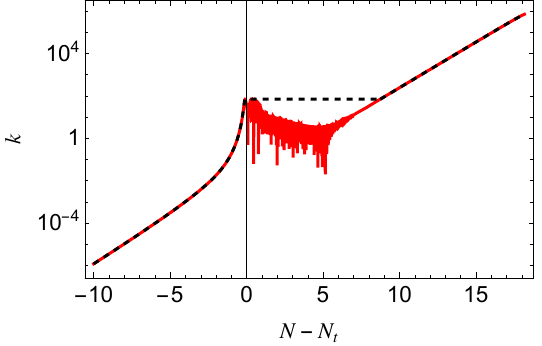}
\caption{Comparison between $|\tilde{k}_{\text{thr}}(N)|$ and $\tilde{k}_{\text{reg}}(N)$ for the set of parameters of fig. \ref{fig:traj}. The black dashed line is the regulator as defined in eq.(\ref{eq:k_regulator}), while the red line is the absolute value of the threshold between stability and instability.
The regulator effectively picks up the maximum value ever reached by $\tilde{k}_{\text{thr}}(N)$.}
\label{fig:regulator_comparison}
\end{figure}

\noindent This function is effectively equal to the maximum value ever reached by $\tilde{k}_{\text{thr}}(N')$ up to $N'=N$. In fig. (\ref{fig:regulator_comparison}), we plot $|\tilde{k}_{\text{thr}}(N)|$ (red line) and $ \tilde{k}_{\text{reg}}(N)$ (dashed black line).

Now, we discuss the UV regulation of the mode functions and the auxiliary function. We want to keep a given mode fixed to its sub-horizon initials conditions until it becomes smaller than the value $\tilde{k}_{\rm{vac}}(N)$ defined in Section \ref{sec:numerical}. In order to achieve this, we employ the same regulating function used for the backreaction integral, with the substitution $\tilde{k}_{\rm{reg}}(N) \rightarrow \tilde{k}_{\rm{vac}}(N) $ and we write the equations of motion as

\begin{equation}
    \frac{d^2 \tilde{A}_\lambda}{d N^2} = R (N,k) \left\{ \lambda \frac{\tilde{k}}{a h} \left( c_F \frac{d \tilde{\theta}}{dN} + c_G \frac{d \tilde{\rho}}{dN} \right) \tilde{A}_\lambda
    - \left\{ \frac{2 \tilde{V}}{h^2} - \frac{1}{6 \tilde{n}_2} \left[ \left( \frac{d \tilde{\theta}}{dN} \right)^2 + \left( \frac{d \tilde{\rho}}{dN} \right)^2 \right] -1 - 2i \frac{\tilde{k}}{a h} \right\} \frac{d \tilde{A}_\lambda}{d N}  \right\}.
\end{equation}
In this way, the given mode $k$ is frozen until $\tilde{k}_{\rm{vac}}(N)$ becomes larger than $k$.
This is made possible by the fact that in the vacuum regime, the first derivative is zero, therefore we only need to set the second derivative to zero to ensure that the mode is frozen. The same redefinition of the equation of motion is performed for the auxiliary function $\tilde{\cal{F}}$ through eq.(\ref{eq:eom_F_tilde}).

\begin{figure}[h]
    \centering
    \begin{subfigure}[t]{0.4\textwidth}
        \centering
        \includegraphics[scale=0.8]{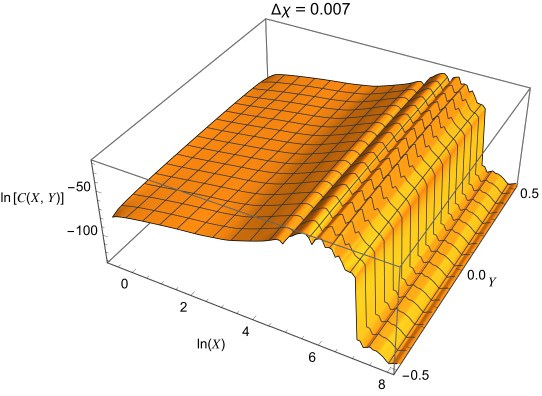} 
    \end{subfigure}
    \hspace{2.7cm}
    \begin{subfigure}[t]{0.4\textwidth}
        \centering
        \includegraphics[scale=0.8]{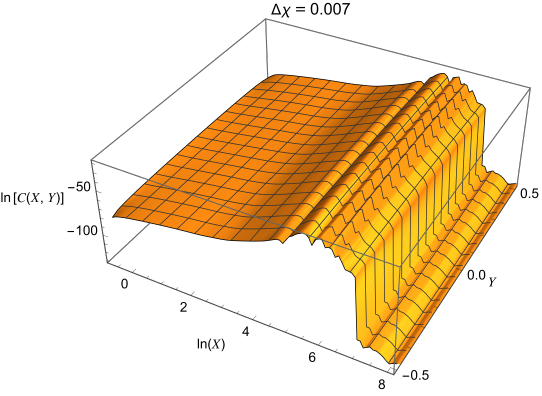}
    \end{subfigure}

    \vspace{1em} 

    \begin{subfigure}[t]{0.4\textwidth}
        \centering
        \includegraphics[scale=0.8]{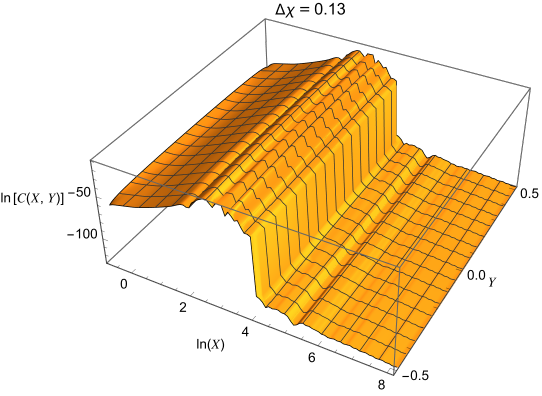} 
    \end{subfigure}
    \hspace{2.7cm}
    \begin{subfigure}[t]{0.4\textwidth}
        \centering
        \includegraphics[scale=0.8]{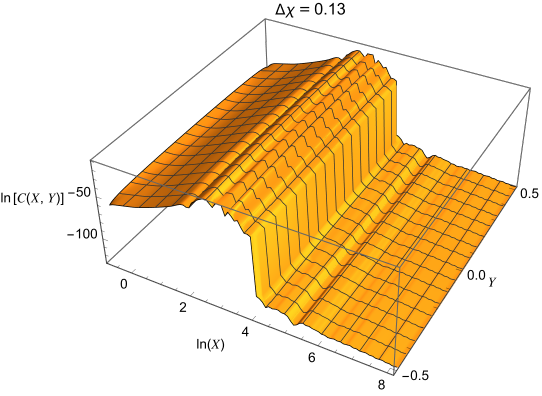} 
    \end{subfigure}
    \vspace{1em} 
    

    \begin{subfigure}[t]{0.4\textwidth}
        \centering
        \includegraphics[scale=0.8]{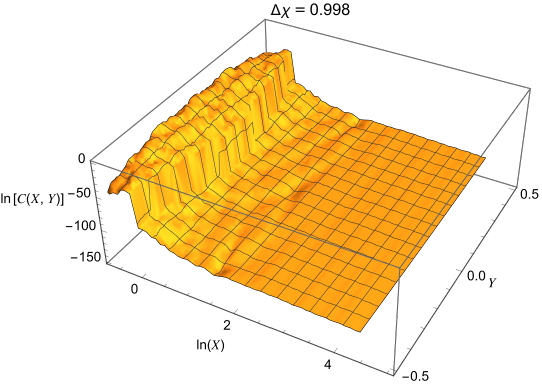}
        \label{fig:image3}
    \end{subfigure}
    \hspace{2.7cm}
    \begin{subfigure}[t]{0.4\textwidth}
        \centering
        \includegraphics[scale=0.8]{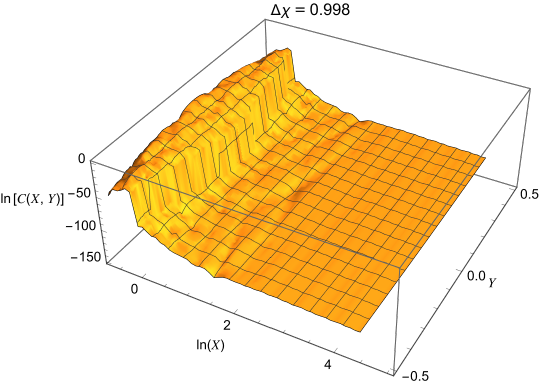} 
    \end{subfigure}

    \caption{Plots of the logarithm of the integrand function ${\cal{C}}(X,Y)$ defined in eq. (\ref{eq:loop_integral}) for different values of $k$ (different rows). The left (right) panels show the left (right)-handed integrand for each $k$.}
    \label{fig:loop_integrand}
\end{figure}

\section{Considerations on the loop integral}
\label{app:loop_integral}

In this appendix, we discuss the loop integral of eq.(\ref{eq:loop_integral}) to exemplify the discussion of Section \ref{sec:results}. A gravitational wave of momentum \(\vec{k}\) is sourced by two gauge modes with momentum $\vec{p}$ and $\vec{q}$ satisfying $\vec{p}+\vec{q}=\vec{k}$. We perform the loop integral in terms of the variables $X,\;Y$ defined in Appendix \ref{App:Code_variables}. In fig. \ref{fig:loop_integrand}, we show the logarithm of the integrand ${\cal{C}}(X,Y)$ defined in Appendix \ref{App:Code_variables} as a function of $\ln(X)$ and $Y$. Every row in the figure shows the integrand for the  left (right) polarization of the GW on the left (right) panel for given value of $\vec{k}$. The three different momenta are respectively $\simeq 0.06,\;1,\;43$ in units of $k_0$ and they correspond to the frequencies marked with the star in fig. \ref{fig:power_spectrum}, from smaller to larger frequencies respectively. Every panel also shows the value of the chirality parameter defined in eq. (\ref{eq:Chirality}).

For $k/k_0 \simeq 0.06$, the integral is dominated by $X \simeq {\cal{O}} (10^2)$. This is because the gauge modes that produce a GW with $k \ll k_0$ have necessarily $\vert p \vert,\;\vert q \vert \gg \vert k \vert$ since the gauge modes with $\vert p \vert,\;\vert q \vert \lesssim \vert k \vert$ are not excited by the transition. Therefore, the two gauge momenta have to be anti-aligned and much greater than $k$, this corresponds to $X \gg Y$ which is indeed what we see in the first row of Fig. \ref{fig:loop_integrand}. In this limit, the polarization factor becomes polarization-independent, as one can see from eq.(\ref{eq:polarization_factor}), therefore, we see a small value of the chirality parameter.

In the second row, we have $k/k_0 \simeq 1$. In this case, the integral is mostly sourced by gauge modes with momentum $\vert p \vert,\;\vert q \vert \gtrsim k$, therefore we have that the integral is still dominated by $X \gg Y$, but the peak is at $X \simeq {\cal{O}}(10)$, therefore we have a small suppression of the right-handed polarization due to the polarization factor. And thus, we see an increase of the chirality of the power spectrum.

In the third row, we have \( k/k_0 \simeq 43\), meaning the power spectrum is dominated by gauge modes with \( |p|, |q| \lesssim |k| \). These momenta must be aligned to sum up to \( k \), a condition satisfied at \( X = 1/2 \). Indeed, we observe that the integral is dominated by \( X \simeq 1/2 \). In this limit, only left-handed photons contribute to left-handed GWs, while only right-handed photons contribute to right-handed GWs. If we neglect the right-handed photons, the right-handed GWs experience strong suppression, leading to a highly chiral power spectrum. This strong suppression is alleviated if we consider the right-handed photon, since it is the only one contributing to the right-handed GWs.
However, as it is still produced less than the left-handed photon, we have only a moderate enhancement (about \( 1 \)–\( 2 \) orders of magnitude) of the right-handed power spectrum and $\Delta \chi$ still remains close to 1.

\section{Analytic study of the second inflationary stage}
\label{App:second-inflation}
To obtain the valley connected to the minimum of the potential, along which the second stage of inflation takes place, we have to find the light direction corresponding to the light eigenmass at the origin of the potential. The valley can be well approximated by by a straight line along this light direction, since the first inflationary stage ends close enough to the minimum of the potential.
The eigenmasses of the potential at the origin are given by the eigenvalues of the hessian, therefore, we find the light and heavy masses respectively $m_l$ and $m_h$
\begin{eqnarray}
    m_l^2 & = &\frac{\Lambda^4}{1+r_\Lambda} \left[ \frac{r_\Lambda {\cal{C}}^2}{n_1+r_\Lambda n_2}+ {\cal{O}}({\cal{C}}^4) \right], \nonumber \\
    m_h^2 &= &\frac{\Lambda^4}{1+r_\Lambda} \left[n_1+ r_\Lambda n_2  - \frac{ r_\Lambda {\cal{C}}^2}{n_1+ r_\Lambda n_2}  + {\cal{O}}({\cal{C}}^4)  \right].
\end{eqnarray}
One can find the two normalized eigenvectors respectively $\vec{v}_l$ and $\vec{v}_h$
\begin{eqnarray}
   v_{h,1}&=&\frac{{\cal{C}} \sqrt{2 n_1 n_2-2{\cal{C}}^2}}{\sqrt{n_2 \left[C^2 \left(2 y-4 n_2 r_\Lambda\right)+n_1^2 n_2+n_1 n_2 \left(2 n_2 r_\Lambda-y\right)+n_2^2 r_\Lambda \left(n_2 r_\Lambda-y\right)\right]}} = - v_{l,2},\nonumber \\
    v_{h,2}&=& - \frac{{\cal{C}} \sqrt{2 n_1 n_2-2{\cal{C}}^2}}{\sqrt{ n_2 \left[ n_1^2 n_2 +n_1 n_2 (2 n_2 r_\Lambda - y) - 2 {\cal{C}}^2 (y+2 n_2 r_\Lambda)+r_\Lambda n_2^2 (r_\Lambda n_2 + y)  \right]}} = -v_{l,1},
    \label{eq:heavy-eigenvector}
\end{eqnarray}

where $y= \sqrt{(n_1+ r_\Lambda n_2)^2 - 4 {\cal{C}}^2 r_\Lambda}$. From the light eigenvector, we can define the trajectory as 
\begin{equation}
    \rho_{\rm{traj}}= \frac{v_{l,2}}{v_{l,1}}(\theta_{\rm{traj}}-\theta_0)+\rho_0,
\end{equation}
where $\{\theta_0, \rho_0 \}$ parameterize the set of trajectories passing for the different, periodic, minima of the potential. If we start from the saddle point of fig. (\ref{fig:traj}), having fixed the gauge with the choice $g_2^{-1}=0$, we have $\theta_0=2 \pi f_2$ and $\rho_0= - \frac{2 \pi g_1 f_2}{f_1}$. 

At this point, we introduce the canonical field $\varphi$ to parameterize the rajectory along the valley
\begin{equation}
    \varphi=\sqrt{1+\left( \frac{d \rho_{\rm{traj}}}{d \theta_{\rm{traj}}} \right)^2} \left( \theta_{\rm{traj}}- \theta_0 \right)= \sqrt{1+\left( \frac{v_{l,2}}{v_{l,1}}\right)^2} \left( \theta_{\rm{traj}}- \theta_0 \right),
\end{equation}
the equation for the valley is therefore expressed in terms of the two parameterized equations
\begin{eqnarray}
    \theta (\varphi)&=& \varphi \left(  1+ \frac{v_{l,2}^2}{v_{l,1}^2} \right)^{-1/2}  + \theta_0, \nonumber \\
    \rho (\varphi)&=&  \frac{v_{l,2}}{v_{l,1}} \varphi \left(  1+ \frac{v_{l,2}^2}{v_{l,1}^2} \right)^{-1/2} + \rho_0.
    \label{eq:parametrized_theta_rho}
\end{eqnarray}
Substituting the two above equations into the potential and expanding to second order in ${\cal{C}}$ and $\varphi$ we find the effective potential of eq.(\ref{eq:Veff}).
In the slow -roll approximation, the equation of motion is given by
\begin{equation}
    \frac{d \varphi}{d N} \simeq -\frac{M_p^2}{V} \frac{d V}{d \varphi} = -2 \frac{M_p^2}{\varphi},
\end{equation}
which can be solved finding
\begin{equation}
    \varphi (N) = \sqrt{\varphi_0^2-4 M_p^2 N}.
    \label{eq:phi(N)}
\end{equation}
Inflation ends when the slow-roll parameter $\epsilon=\frac{M_p^2}{2} \left( \frac{1}{V} \frac{d V}{d \varphi} \right)^2$ is equal to 1, which corresponds to \\
$\varphi_{\rm{end}}=\sqrt{2} M_p$. Therefore, substituting into eq.(\ref{eq:phi(N)}) and solving for $N$, we find the total number of e-folds
\begin{equation}
    N_{\rm{tot}}= \frac{\varphi_0^2}{4 M_p^2} - \frac{1}{2}.
\end{equation}
At this point, all we have to do is to estimate $\varphi_0$ from the parameters of the model, and we do so exploiting eq.(\ref{eq:phi0}), where $\rho_{\rm{end}}$ is given by  
\begin{equation}
\label{eq:rho_end_approx}
    \rho_{\rm end} = - \frac{1}{f_1 \, {\cal C}} \, \arccos \left( \frac{1}{r_\Lambda} \, \sqrt{\frac{n_1 \left( n_1 - r_\Lambda^2 \, n_2 \right)}{n_2 \left( n_2 - n_1 \right)}} \right) + \frac{1}{f_2 \, {\cal C}} \, \arcsin \left( \sqrt{\frac{r_\Lambda^2 n_2^2 - n_1^2}{n_1 \left( n_2 - n_1 \right)}} \right) + \frac{f_2^{-1} \, \pi}{f_1^{-1} g_2^{-1} - g_1^{-1}  f_2^{-1}}.
\end{equation}

\section*{Acknowledgments}
I am grateful to Marco Peloso for his careful reading of the manuscript and for the valuable discussions and suggestions. I also wish to thank Alexandros Papageorgiou for generously sharing the \textit{Mathematica} code that has later been extendend to obtain the results of this work. I acknowledge support from Istituto
Nazionale di Fisica Nucleare (INFN) through the Theoretical Astroparticle Physics (TAsP) project and from Dipartimento di Fisica e Astronomia (DFA) of the University of Padua.

\bibliographystyle{jhep}
\bibliography{bibliography}

\end{document}